\documentclass[11pt]{article}

\usepackage{a4}

\usepackage{amsmath}
\usepackage{graphicx}
\usepackage{latexsym}
\usepackage{amsfonts}
\usepackage{amssymb}
\usepackage{epsfig}
\usepackage{psfrag}
\usepackage{dsfont}

\usepackage[sc,small]{caption2}
\setcaptionwidth{14cm}

\makeatletter
\@addtoreset{equation}{section}
\makeatother

\newcommand{\io}{\iota^*}
\newcommand{\F}{\mathcal F}
\newcommand{\V}{\mathcal V}
\newcommand{\non}{\nonumber \\}

\newcommand{\cf}{{\cal F}}
\newcommand{\tr}{{\rm tr}}

\newcommand{\ff}{\ensuremath{\mathbb{F}}}
\newcommand{\cc}{\ensuremath{\mathbb{G}}}
\def\Re{\mathop{\rm Re}\nolimits}
\def\Im{\mathop{\rm Im}\nolimits}
\begin{document}

%%%%%%%%%%%%%%%%%%%%%%%%%%%%%%%%%%%%%%%%%%%%%%%%%%

\begin{titlepage}
\begin{flushright}
hep-th/0609211\\
LMU-ASC 64/06\\
MPP-2006-123\\
KUL-TF-06/24\\
\end{flushright}
\vspace{.1cm}
\begin{center}
\baselineskip=16pt {\bf \LARGE Gaugino Condensates and \\[3mm] $D$-terms from D7-branes}\\
\vspace*{7mm} \vfill%17.mm
{\Large
Michael Haack ${}^a$, Daniel Krefl ${}^a$, Dieter L{\"u}st ${}^{a,b}$, \\ \vskip2mm
Antoine Van Proeyen ${}^c$ and Marco Zagermann ${}^b$
} \\
\vspace*{3mm} \vfill%0.5cm
{\small
${}^a$ Arnold Sommerfeld Center for Theoretical Physics \\
Department f{\"u}r Physik, Ludwig-Maximilians-Universit{\"a}t M{\"u}nchen,  \\
Theresienstra{\ss}e 37, 80333 Munich, Germany\\ \vskip2mm
${}^b$ Max-Planck-Institut f{\"u}r Physik, F{\"o}hringer Ring 6,\\
80805 Munich,  Germany\\ \vskip2mm
${}^c$ Instituut voor Theoretische Fysica, Katholieke Universiteit Leuven\\
Celestijnenlaan 200D, B-3001 Leuven, Belgium \\ \vskip4mm  }
\end{center}
\vfill
\begin{center}
{\bf Abstract}
\end{center}
{We investigate, at the microscopic
level, the compatibility between $D$-term potentials from
world-volume fluxes on D7-branes and
non-perturbative superpotentials arising from
gaugino condensation on a different
stack of D7-branes. This is motivated by attempts
to construct metastable de Sitter vacua in type IIB string theory
via $D$-term uplifts. We find a condition under which the
K\"ahler modulus, $T$, of a Calabi-Yau 4-cycle 
gets charged under the anomalous $U(1)$ on the
branes with flux. If in addition this 4-cycle is wrapped 
by a stack of D7-branes on which gaugino condensation takes place,
the question of $U(1)$-gauge invariance of the
($T$-dependent) non-perturbative superpotential arises.
In this case an index theorem guarantees that 
strings, stretching between the two stacks,
yield additional charged chiral fields which also
appear in the superpotential from gaugino condensation.
We check that the charges work out to make this
superpotential gauge invariant, and we argue that the
mechanism survives the inclusion of higher curvature corrections
to the D7-brane action. 
}\vspace{2mm} \vfill \hrule width 3.cm
{\footnotesize \noindent e-mail: \{haack,krefl,luest\}@theorie.physik.uni-muenchen.de \\
\hspace*{1.1cm}  antoine.vanproeyen@fys.kuleuven.be \\
\hspace*{1.1cm}  zagerman@mppmu.mpg.de
\\
\noindent }
\end{titlepage}
\tableofcontents{}
%\newpage
%%%%%%%%%%%%%%%%%%%%%%%%%%%%%%%%%%%%%%%%%%%%%%%%%%%%%

\section{Introduction}

One of the most far-reaching consequences of the recent impressive advances in
observational cosmology is the strong evidence for an accelerated expansion
of the present universe, apparently driven by a small positive cosmological
constant, or a similar form of ``dark energy''.

Implementing  a positive cosmological constant in a semi-realistic string
theory set-up has been a subject of great interest during the past
couple of years. In the standard low energy framework of string compactifications,
one attempts to find local positive minima of the effective
4D supergravity potential in which all moduli are stabilized at sufficiently
large mass scales so as to respect several other phenomenological constraints.
Whereas moduli stabilization in general is already
inflicted with a number of technical challenges (many of which have been overcome
during the past years), the stabilization of moduli in
a local de Sitter (dS) minimum requires some extra care due to a number of
peculiarities of de Sitter spacetimes.

Some of these problems can already be understood from the lower-dimensional
supergravity point of view and are due to the fact that  there is no dS  analogue of the Anti-de Sitter (AdS)
Breitenlohner-Freedman bound \cite{Breitenlohner:1982jf}
and   that supersymmetry is necessarily broken in a dS
background.   The first property
implies that, unlike for some AdS spaces, the stability of a dS vacuum does not tolerate
tachyonic directions at the critical point, whereas the lack of supersymmetry
means that tachyonic
directions cannot  be ruled out using any supersymmetry arguments, as is e.g.
possible for supersymmetric Minkowski vacua  (for a recent  discussion, see
\cite{Kallosh:2004yh,Blanco-Pillado:2005fn,Krefl:2006vu}).\footnote{If a
dS vacuum arises from a ``gentle'' uplift
of a supersymmetric Minkowski vacuum, as
in \cite{Kallosh:2004yh,Blanco-Pillado:2005fn,Krefl:2006vu}, one might
obtain a nearby dS solution that inherits the absence of tachyonic
directions from the Minkowski vacuum.} These observations alone   make
stable dS vacua in supergravity theories appear  less generic than stable
Minkowski or AdS vacua.\footnote{In fact,  despite the enormous body of
literature from the 1980's, stable de Sitter vacua were not found in
extended gauged supergravity theories until rather
recently \cite{Gunaydin:2000xk,Fre:2002pd,Fre':2003gd,Cosemans:2005sj}.}

Besides these purely lower-dimensional issues, there are additional
difficulties if one wants to obtain
dS vacua from a compactification of 10D/11D    string/M-theory.
   In fact,
there exist several  no-go theorems against dS compactifications
\cite{Gibbons,dWSD,Maldacena:2000mw},
and finding a consistent dS compactification requires a violation of
at least one of the premises
that underlie these no-go theorems.

In \cite{Kachru:2003aw},  a strategy for the construction of (meta-)stable
dS vacua in a class of type IIB compactifications was proposed,
based on a  combination and a careful balance of a number of different
effects. The key ingredients used in \cite{Kachru:2003aw} are
   background fluxes on a type IIB  $CY_{3}$ orientifold  (or an F-theory
generalization thereof), a non-perturbative superpotential, $W_{\rm np}$, due to
gaugino condensation on wrapped D7-branes or Euclidean D3 instantons, as well as
a contribution to the scalar potential due to
anti-D3-branes ($\overline{D3}$-branes). The r\^{o}le of the background
fluxes is to generate a superpotential $W_{\rm flux}$ for the complex structure
moduli and the dilaton, whereas the non-perturbative superpotential $W_{\rm np}$
leads to a non-trivial potential also for the K{\"a}hler moduli.

The gaugino condensation of pure SYM leads to a non-perturbative
superpotential of the  form
  \begin{equation}
 W_{\rm np} = A(z) e^{-\alpha T^{\cc}} ,\label{Wsimple}
 \end{equation}
 where $T^{\cc}$ is the complex  K{\"a}hler modulus of the 4-cycle
  wrapped by the D7-branes on which the gaugino condensation takes place,
and $A(z)$ denotes a possible function of the complex structure moduli, the
dilaton and the open string fields, which arises due to 1-loop threshold
corrections to the D7-brane gauge kinetic
function and/or due to world-volume fluxes and higher curvature corrections
  on the world volume of the
D7-branes.\footnote{The superpotential in the case of Euclidean D3-branes
has a similar form. In this text, we restrict our attention to the mechanism
of gaugino condensation. For a recent discussion of Euclidean
D3-branes, see \cite{Ralph,Ibanez:2006da}.
In writing down (\ref{Wsimple}) we assumed
that there is no light matter charged under the D7-brane gauge group. We will come back to
the more general case including matter in the (anti)fundamental
representation below.}  If, as in the original proposal  \cite{Kachru:2003aw},
one has only one K{\"a}hler modulus denoted by $T$,
 $W_{\rm flux} +W_{\rm np}$ can
generate AdS minima
in which all moduli are stabilized. These AdS minima can  finally be
``uplifted'' by the $\overline{D3}$-brane contribution to the potential,
\begin{equation}
  V_{\overline{D3}}\sim \frac{\mu_{3}}{[\Re (T)]^2}\ ,   \label{barD3}
\end{equation}
where $\mu_{3}$ is the tension of the $\overline{D3}$-brane.
This may lead to a stable dS vacuum under certain conditions
\cite{Choi:2004sx,Lust:2005dy,Lust:2006zg}.

In a variant of this construction,
Burgess, Kallosh and Quevedo \cite{Burgess:2003ic} suggested an alternative
mechanism for the uplift of the AdS to the dS minimum that is not based on a
$\overline{D3}$-brane, but
instead on world volume fluxes on  D7-branes. In contrast to the
$\overline{D3}$-brane potential, the effects of the D7-brane world volume
fluxes can easily be incorporated in the framework of
a standard supergravity Lagrangian, which helps one to maintain
technical control at  all  stages of the 
construction.\footnote{Similar uplifting methods using non-perturbative 
potentials in the context of the heterotic 
string were discussed in \cite{Becker:2004gw,Buchbinder:2004im,Curio:2005ew,Curio:2006dc}.}

In order to understand the effect of the world volume fluxes, let us
  denote by D7$_\cc$    the D7-brane stack on which  the gaugino condensation
  takes place  and by D7$_\ff$  a  D7-brane
(stack) with a non-trivial  world-volume flux. A priori, these two stacks could
be the same and/or wrap the same 4-cycle, which is the original situation described
in \cite{Burgess:2003ic}. This case was already discussed in \cite{Jockers:2005zy}
and we will, in this paper, always assume that
D7$_{\cc}$ and D7$_{\ff}$ are different and wrap different 4-cycles.
Likewise, the   K{\"a}hler moduli of the 4-cycles $\Sigma^{\ff}$ and $\Sigma^{\cc}$
that are   wrapped by
D7$_\ff$ and D7$_\cc$ are denoted by $T^{\ff}$ and $T^{\cc}$, respectively, 
as already done in (\ref{Wsimple}), and
we use $G_\ff$ and $G_\cc$ for the respective  gauge groups realized on these
brane stacks. $G_\ff$ is assumed to have an Abelian factor $U(1)_\ff$.
A generic four-cycle will be denoted by $\Sigma^{\alpha}$.

Under certain topological conditions to be derived in the next section,
the effect of the world volume flux on D7$_\ff$ is to generate a $D$-term that corresponds
to a gauged shift symmetry of the  K{\"a}hler  modulus $T^\cc$,
\begin{equation}
T^\cc \rightarrow  T^\cc+iq\epsilon,  \label{TTq}
\end{equation}
with ``charge'' $q$ and gauge parameter $\epsilon$,
so that the axion $a^\cc      \equiv    \Im(T^\cc)$ appears with a gauge
covariant derivative
\begin{equation}\label{Lag}
\mathcal{L} \sim (\partial_{\mu} a^\cc  + qA^{(\ff)}_{\mu})^2
\end{equation}
in the effective supergravity Lagrangian.\footnote{To
anticipate the result,
a necessary condition for the appearance of
a non-vanishing $D$-term involving $T^{\cc}$
is that the 4-cycles
wrapped by D7$_{\cc}$ and D7$_{\ff}$ intersect over a 2-cycle on which the
world-volume flux is non-trivial, cf.\ (\ref{Q}) and (\ref{DBRANEeq30}).}
In this expression,
  the vector field $A^{(\ff)}_{\mu}$ arises from  the $U(1)_{\ff}$
  world volume gauge field on
D7$_{\ff}$.
The gauging of (\ref{TTq}) entails  a non-trivial $D$-term potential,
\begin{equation}
V_{D} = \frac{1}{2 g^{-2}_{\ff}} \Big[ M_{\rm P}^2 q \partial_{T^\cc}  K  + \sum
M_{\rm P}^2 q_{I} |\Phi_I|^2 \Big]^{2} , \label{Dterm}
\end{equation}
where we introduced the reduced Planck mass $M_{\rm P}^2$ (which appears in the
Einstein-Hilbert action as $\frac{M_{\rm P}^2}{2} \int d^4x \sqrt{-g} R$).\footnote{For the
appearance of the $M_{\rm P}^2$-factors, compare with formulas (4.6) and (5.15) in
\cite{Kallosh:2000ve}.  In general, there will be other K{\"a}hler moduli charged under $U(1)_{\ff}$, cf.~eq.~(\ref{dD7}), which we did not display in (\ref{Dterm}).} The first term in  (\ref{Dterm}), $q\partial_{T^{\cc}} K$,
arises from the standard expression $D\sim \eta^{T^\cc} \partial_{T^{\cc}}  K $
for a $D$-term with  K{\"a}hler potential $K$ and  a constant
Killing vector $\eta^{T^{\cc}}=iq$ corresponding to the gauged shift symmetry
(\ref{TTq}). The fields $\Phi_{I}$ in (\ref{Dterm})
denote other possibly present fields which transform linearly under
$U(1)_{\ff}$ with charges $q_{I}$ and which, for simplicity,
are assumed to have a canonical K{\"a}hler potential.
The tree-level gauge kinetic function yields
$g_{\ff}^{-2}\sim \Re(T^\ff) + \ldots$, where
the dots are flux-dependent and/or curvature  corrections   involving the dilaton (see eq. (\ref{fff})).
In some simple cases, the tree-level K{\"a}hler potential  for $T^{\cc}$ is of the form
$K(T^{\cc}+\bar T^{\cc})\sim \ln(T^{\cc}+\bar T^{\cc})$.
To make contact with \cite{Burgess:2003ic}, we
assume this simple K{\"a}hler potential for the moment,
neglect the above-mentioned (or any other)  corrections
to the gauge kinetic function and specialize to the case of one K{\"a}hler modulus,
denoted by $T$.
Furthermore,
assuming that any additional matter field $\Phi_{I}$ has a
vanishing vacuum expectation value (vev), one finds that
the above $D$-term is simply proportional to
$(\Re(T))^{-3}$ and can be used to uplift an AdS minimum,
just as the $\overline{D3}$ potential (\ref{barD3}) \cite{Burgess:2003ic}.

The question as to whether the vevs
of the $\Phi_I$  can really be assumed to be zero is discussed e.g.\ in
\cite{Binetruy:1996uv,Burgess:2003ic,Achucarro:2006zf} and is model-dependent.
As was argued in \cite{Achucarro:2006zf}, the $\Phi_{I}$ are in fact often
 non-zero at the minimum, and the real question is instead
whether the $\Phi_{I}$ can  acquire  vevs that lead to a vanishing
$D$-term or whether $V_{D}\neq 0$ at the minimum.\footnote{In \cite{Lebedev:2006qq},
by contrast,
the $\Phi_{I}$ play an important role in the uplifting, as they
were assumed to give the dominant contribution
to the ($F$-term) potential in the vacuum.}
In any case, as stressed in \cite{Choi:2005ge,Villadoro:2005yq},
one should keep in mind that
a $D$-term potential can only be used to uplift a \emph{non}-supersymmetric AdS $F$-Term vacuum.

While this $D$-term uplifting mechanism looks quite compelling,  it was
already pointed out in  \cite{Burgess:2003ic} and further stressed in
\cite{Binetruy:2004hh,Villadoro:2005yq,Achucarro:2006zf,Parameswaran:2006jh,Dudas:2006vc}
 that  there seems to be a tension between the necessity of the
gauging of the shift symmetry (\ref{TTq}) %$T^\cc \rightarrow T^\cc+iq\epsilon $
and the fact that the
 sum $W_{\rm flux}+W_{\rm np}$ of the superpotentials
  appears non-invariant
under this shift. In fact, there is another seemingly  non-invariant term,
 \begin{equation}
 \Im(T^\cc) \textrm{tr}[F^\cc\wedge F^\cc],   \label{TFF}
 \end{equation}
 where $F^\cc$ denotes the Yang-Mills field of the gauge group $G_\cc$,
  whose
gauginos condense,
 and the prefactor arises from the tree-level relation $f^{\cc} \sim T^\cc +\ldots $
for the   corresponding  gauge
kinetic function. Both non-invariances can be cured by the same mechanism
  if bifundamental matter fields  with non-vanishing mixed anomalies
of the type $U(1)_\ff - [G_\cc]^2$  are present
  \cite{Achucarro:2006zf}. If this is the case, the mixed anomaly can cancel the classical
 non-invariance of (\ref{TFF}) via the Green-Schwarz mechanism,
 and the non-perturbative superpotential   (\ref{Wsimple}) goes over to the
Affleck-Dine-Seiberg (ADS) superpotential
\begin{equation} \label{wads}
 W_{ADS} = \left( \frac{e^{- 8 \pi^2 f^{\cc}}}{\det(M)} \right)^{\frac{1}{N_\cc-N_F}}
= A(z) \left( \frac{e^{- 8 \pi^2 T^{\cc}}}{\det(M)} \right)^{\frac{1}{N_\cc-N_F}}\ ,
\end{equation}
which now contains   an
 additional factor involving the meson determinant \cite{ADS,TVY}
 \begin{equation}
 \det(M^{i}_{j})\equiv \det (\tilde{\Phi}^{ia}\Phi_{ja}),
 \end{equation}
 where $i,j$ denotes the flavor and $a,b$ the color index. The ``quarks'' and ``anti-quarks''
$\Phi_{jb}$, $\tilde{\Phi}^{ia}$ are also charged under the $U(1)_\ff$ vector field
   $A^{(\ff)}_{\mu}$ and hence form part of the charged fields $\Phi_{I}$
in eq. (\ref{Dterm}).
Obviously,  the net sum of the $U(1)_\ff$ charges of the ``quarks''
and ``anti-quarks'' equals the
total $U(1)_\ff$ charge of the determinant $\det(M^{i}_{j})$. Thus, if this total charge
is non-zero, the non-invariance of the $e^{-\alpha T^\cc}$-factor in $W_{\rm np}$ can 
in principle be compensated
by the meson determinant, if the charges work out correctly. 
In this case, the mixed triangle anomaly is also non-vanishing,
and the Green-Schwarz mechanism can operate to cancel the non-invariance of
(\ref{TFF}),
so that the whole mechanism appears automatically self-consistent.

This idea was put forward and discussed at
a field theoretical level in \cite{Achucarro:2006zf,Dudas:2006vc}
(see also \cite{Binetruy:1996uv} for an earlier, related discussion). Here we investigate more concretely a microscopic
implementation of the idea in a D-brane setup, where the charged matter arises
naturally via open strings stretched between the D7$_{\cc}$ and D7$_{\ff}$ stacks.
We verify in particular that the flux-induced charge of the gauged axionic
modulus $a^{\cc}$ and the number of chiral matter fields (charged under both,
$G_{\cc}$ and the anomalous $U(1)_{\ff}$) is precisely such that
the non-perturbative superpotential becomes gauge invariant under $U(1)_{\ff}$
transformations. In this way we confirm the field theoretic ideas of
\cite{Achucarro:2006zf,Dudas:2006vc} within a D-brane setup.

This is a further step in illuminating the proposal of \cite{Burgess:2003ic}.
Our analysis focuses on the issue of the gauge invariance 
of the non-perturbative
superpotential (\ref{wads}) when the shift 
symmetry of the K{\"a}hler modulus $T^{\cc}$
is gauged. More work is required to find a phenomenologically viable
global D-brane model and some of the open issues will be briefly mentioned later on.
Some of the presented ideas are already scattered in one or the other form
in the literature, but we believe that
our composition and elaboration of the facts relevant for embedding
the proposal of \cite{Burgess:2003ic} in a concrete D7-brane model could be helpful
as a stepping stone for building a more complete model.

The organization of this paper is as follows:
In Section 2, we derive the $D$-term due to a  world volume flux on  a D7-brane (D7$_{\ff}$)
wrapped around
a 4-cycle in a general Calabi-Yau using the
Dirac-Born-Infeld action. We observe that a K{\"a}hler modulus gets charged
only if the 4-cycle whose volume the K{\"a}hler modulus measures and the
4-cycle wrapped by the D7-brane intersect over a 2-cycle on which the
world-volume flux is non-trivial. If a 4-cycle whose K{\"a}hler 
modulus is charged,
is also wrapped by some D7-brane(s) (cf.\ the D7$_{\cc}$ 
stack discussed before),
additional chiral bifundamental matter
naturally arises via open strings, exactly as needed for
gauge invariance.\footnote{A related observation was made in
\cite{Jockers:2005zy}, restricting though to chiral matter arising at the
intersection of D7$_{\ff}$ and its orientifold image. This is
not the relevant case if D7$_{\cc}$ and D7$_{\ff}$ are different.}
Variants of the reduction method we apply appeared
in different contexts before, e.g.\ in
\cite{Cremades:2002te,Blumenhagen:2002wn,Blumenhagen:2003vr,Villadoro:2005yq,Martucci:2006ij}. For instance, the
$D$-term potential resulting from world-volume fluxes on the D7-branes
was previously derived in \cite{Jockers:2005zy},
however, with a completely different method,
i.e.\ by analyzing the fermionic terms on the world-volume of the D7-branes.
In our derivation we determine the $D$-term potential 
directly from the bosonic DBI-action and 
keep track of all factors explicitly, which allows us
to read off the charge of the K{\"a}hler modulus $T^{\cc}$ under the anomalous
$U(1)_{\ff}$. This is a prerequisite to verify the gauge invariance of (\ref{wads})
in section 3. Another ingredient in this verification is the number of chiral
(anti)fundamentals of $G_{\cc}$, charged under $U(1)_{\ff}$. This number can be
obtained using the index of the relevant Dirac operator, as we will elaborate on
in section 3. Furthermore, in section \ref{R2} we investigate the effect of
higher curvature corrections to the D7-brane action. We find
that the gauge coupling potentially gets a correction depending on the dilaton
field, but the charge of the K\"ahler moduli is not modified.
Finally, section 4 summarizes our results and concludes.

Appendix \ref{example} illustrates the derivation of the $D$-term potential with
the example of the $\mathbb Z_2\times\mathbb Z_2$ toroidal orientifold in IIB.
Appendix \ref{alternative} gives a further alternative derivation of the $D$-term potential
and appendix \ref{gaugekin} contains the derivation of the imaginary part of the
gauge kinetic function in the D7-brane compactification, including higher curvature
effects.

%%%%%%%%%%%%%%%%%%%%%%%%%%%%%%%%%%%%%%%%%%%%%%%%%%%%%%%%%%

\section{D7-branes in IIB on a Calabi-Yau three-fold}
\label{section:dterm}

%%%%%%%%%%%%%%%%%%%%%%%%%%%%%%%%%%%%%%%%%%%%%%%%%%%%%%%%%%%

In the introduction, we sketched the mechanism proposed  in
\cite{Achucarro:2006zf} that reconciles the presence of a gaugino condensation potential with the $D$-term uplifting mechanism  suggested in \cite{Burgess:2003ic}. In this section, we focus on the $D$-term part of  this set-up by studying   concrete  models with D7-branes and by
deriving in a more precise manner  the resulting $D$-term potential in a KK
reduction along similar lines as
\cite{Cremades:2002te,Blumenhagen:2003vr,Jockers:2005zy,Villadoro:2006ia}.

%%%%%%%%%%%%%%%%%%%%%%%%%%%%%%%%%%%%%%%%%%%%%%%%%%%%%%%%%%%%%%%%%%%%%%%%%%%%%%%%

\subsection{The setup}

%%%%%%%%%%%%%%%%%%%%%%%%%%%%%%%%%%%%%%%%%%%%%%%%%%%%%%%%%%%%%%%%%%%%%%%%%%%%%%%%

We are interested in type IIB orientifolds with $\mathcal N =1$
supersymmetry including D7- and D3-branes.
This can be obtained if the internal Calabi-Yau manifold admits a holomorphic 
involution $\sigma$
that acts on the K{\"a}hler form $J$ and the $(3,0)$-form $\Omega$ of the
Calabi-Yau according to
\begin{equation}
\sigma J = J \quad , \quad \sigma \Omega = - \Omega \ .
\end{equation}
Modding out by
\begin{equation} \label{orientproj2}
\mathcal O = (-1)^{F_L} \Omega_{\rm p} \sigma\ ,
\end{equation}
where $F_L$ is the left moving fermion number and $\Omega_{\rm p}$ 
is the worldsheet parity, leads to O3- and O7-planes at the fixed loci of
$\sigma$, cf.\ \cite{Acharya:2002ag}. This situation was also investigated in 
\cite{Jockers:2004yj,Jockers:2005zy}.
Note that the cohomology classes $H^{(p,q)}$ split into $\sigma$ eigenspaces 
$H^{(p,q)}_+$ and
$H^{(p,q)}_-$. We do not specify the Calabi-Yau further but note that it could 
for example be taken to be
a (blown up) toroidal orbifold $\tilde Y=T^6/\Gamma$, with $\Gamma$
a discrete group of the kind $\mathbb{Z}_N$ or $\mathbb{Z}_N \times \mathbb{Z}_M$.
A systematic analysis
of this class of models has been started in 
\cite{Lust:2004fi,Lust:2005dy,Lust:2006zg,Lust:2006zh}. The Calabi-Yau after 
modding out the involution $\mathcal O$ will be denoted by $Y=\tilde Y / \mathcal O$ in the 
following.

%In the introduction we considered a very special situation by picking  a particular
%D7-brane stack D7$_\cc$,
%on which gaugino condensation  is supposed to take  place,
%and   by  selecting   a --  possibly different --  stack,
%D7$_{\ff}$, carrying a non-trivial world volume flux.
In this section,
we first compute
the contribution of a  single   D7-brane
with  world volume flux to the $D$-term potential, and later take into account the
contribution of other D7/D3-branes (including the mirror branes) and the
O7/O3-planes.
Just as in the introduction, we will use the subscript $\ff$ to denote
all quantities pertaining to this D7-brane with flux. The stack on which
gaugino condensation takes place will likewise be denoted by a subscript
$\cc$; however, it will not enter until the end of section \ref{dbidterm}.

Let us therefore consider  a single D7-brane, which we assume to be part of
 the $\ff$ stack.  To lowest order, its contribution to the effective action
is governed by the Dirac-Born-Infeld (DBI) and Chern-Simons (CS) action
\cite{Polchinski:1998rr}. The DBI action for a D7-brane in the string frame is given by
\begin{equation}\label{DBRANEeq1}
S_{DBI}=-\mu_7\int_{\mathcal W}d^{8}\xi e^{-\phi}\sqrt{-\det(\iota^*g
+\iota^*B+2\pi\alpha' F)}\ ,
\end{equation}
where $\mathcal W$ is the eight-dimensional D7-brane world-volume which splits
into the four-di\-mensio\-nal spacetime part, $\mathcal M_4$, and a
four-dimensional
internal part on a 4-cycle, $\Sigma^\ff$: $\mathcal W=\mathcal M_4\times\Sigma^\ff$.
Furthermore, $\phi$ denotes the ten-dimensional dilaton, $\mu_7$ is the
D7-brane tension,\footnote{Note that this value for the tension is
the appropriate one for T-duals of type I, where the physics is locally oriented
\cite{Polchinski:1998rr}. We use the conventions of \cite{Polchinski:1998rr}
throughout the paper.}
\begin{equation} \label{mu7}
\mu_7= (2\pi)^{-7}\alpha^{\prime -4}\ ,
\end{equation}
$\iota^*g$ and $\iota^*B$, respectively, 
denote the pullbacks of the ten-dimensional metric and the NSNS 2-form
to $\mathcal W$,
%and $\mathcal F$ is the combined 2-form flux defined by
%\begin{equation}
%\mathcal F=\iota^*B+2\pi\alpha' F,
%\end{equation}
%with $\iota^*B$ the pullback of the NSNS 2-form
and $F$ is the
field-strength of the
D7-brane gauge field.
%Since $\iota^*B$ needs to be expanded in $H_{-}^{(p,q)}(\Sigma)$, but
%$H_{-}^{(p,q)}(\Sigma)=\emptyset$ as $\Sigma$ coincides with the
%world-volume of the O-plane, one concludes that $\iota^*B$ vanishes. Hence,
%\begin{equation}
%\mathcal F=2\pi\alpha' F.
%\end{equation}
%
%
The CS action is given by
\begin{equation}\label{DBRANEeq2}
S_{\rm CS}=-\mu_7\int_{\mathcal W}\sum_p \iota^*C_p\wedge e^{(\iota^*B+2\pi\alpha' F)} ,
\end{equation}
where $\iota^*C_p$ denotes the pullback of the respective RR form.

The BPS calibration conditions for D-branes wrapping Calabi-Yau
cycles were derived
in \cite{Marino:1999af}.\footnote{Note that these conditions
also hold under the
inclusion of background fluxes \cite{Gomis:2005wc}.} In detail,
for a D7-brane,
the wrapped 4-cycle $\Sigma^{\ff}$ needs to be a divisor and has to be
holomorphically
embedded into the Calabi-Yau.
%Equivalently, the following relations need to hold on $\Sigma$ \footnote{Note that we use
%$\mathcal F$ to denote the full 2-form on $\mathcal W$ as well as
%the internal 2-form flux on $\Sigma$. The respective meaning
%should be clear from the context. }
This can be equivalently expressed as
\cite{Marino:1999af}:
\begin{equation}\label{DBRANEeq3}
\frac{1}{2}\left(\iota^*J+i\mathcal F\right)\wedge\left(\iota^*J+i\mathcal
F\right)=e^{i\theta}\sqrt{\frac{\det(g_{\Sigma^{\ff}}+\mathcal F)}{\det{(g_{\Sigma^{\ff}})}}}
\textrm{Vol}_{\Sigma^{\ff}}\ ,
\end{equation}
\begin{equation}\label{DBRANEeq4}
\mathcal F^{2,0}=\mathcal F^{0,2}=0\ ,
\end{equation}
where $\theta$ is an a priori arbitrary phase,
%$g_\Sigma$ the pullback of $g$ to
%$\Sigma$ (i.e.\ $g_\Sigma=(\io g)|_\Sigma$)
$\textrm{Vol}_{\Sigma^{\ff}}$ denotes the volume form on $\Sigma^{\ff}$, and we have used\footnote{In the rest of this paper, we will sometimes also use the symbol $\mathcal{F}$ for the full  expression $(\iota^*B+2\pi\alpha' F)$, i.e.,
not only for the restriction to $\Sigma_{\ff}$. It should be clear from the context which of these two meanings of $\mathcal{F}$ is intended.}
\begin{eqnarray}
g_{\Sigma^{\ff}}&=&(\io g)|_{\Sigma^{\ff}} \\
\mathcal F&=&(\iota^*B+2\pi\alpha' F)|_{\Sigma^{\ff}}.
\end{eqnarray}
Moreover,
$\iota^*J$ is the pullback of the K{\"a}hler form which we can expand into
harmonic forms $\iota^*\omega_\alpha$ of $H^{(1,1)}(\Sigma^{\ff})$:
\begin{equation}\label{DBRANEeq5}
\iota^*J=v^\alpha\iota^*\omega_\alpha\ ,
\end{equation}
where $\omega_\alpha$ form a basis of $H^{(1,1)}(Y,\mathbb{Z})$.

In writing down (\ref{DBRANEeq1}), (\ref{DBRANEeq2}) and (\ref{DBRANEeq3}),
we neglected higher derivative corrections
involving the Riemann tensor, cf.\
\cite{Green:1996dd,Dasgupta:1997cd,Cheung:1997az,Minasian:1997mm,Bachas:1999um,Fotopoulos:2001pt}.
For the moment we just
assume that they do not contribute, but we will come back to them in section \ref{R2}.

In the following, we assume, for simplicity, that
the negative $\sigma$ eigenspaces $H_{-}^{(q,2-q)}(\tilde Y)$  vanish:
\begin{equation}
H_{-}^{(q,2-q)}(\tilde Y)=0.
\end{equation}
This implies that the NSNS 2-form $B$, which is odd under world-sheet parity
and hence, as it lives in the whole bulk,  needs to be expanded in elements of the above negative cohomology groups, vanishes.
Otherwise, the definition of the K{\"a}hler moduli becomes rather
cumbersome   \cite{Jockers:2005zy}.\footnote{The additional moduli arising from $B$ in the case of $H_{-}^{2}(\tilde Y)\neq 0$ can be stabilized by $D$-terms \cite{Lust:2006zg}.}
This also means that $B$
does not contribute to $\mathcal{F}\equiv(\iota^*B+2\pi\alpha' F)|_{\Sigma^{\ff}}$, i.e., we assume
\begin{equation}
\mathcal F=2\pi\alpha' F.
\end{equation}
This would automatically be true if the D7-brane was sitting on top of an O7-plane,
as $\Sigma^{\ff}$ would then also be wrapped by the O7-plane, i.e.\ it would be $\sigma$-invariant.
However, we do not assume that the D7-brane is sitting on top of the O7-plane
here. We will come back to a more precise description of the D-brane setup
at the beginning of section \ref{anomalies}.

In general, the forms $\iota^*\omega_\alpha$ which arise as pullbacks of $(1,1)$-forms
of the ambient Calabi-Yau $Y$, do not necessarily form a
complete basis of $H^{(1,1)}(\Sigma^{\ff})$ \cite{Jockers:2005zy}. There might be
additional $(1,1)$-forms that are harmonic only locally on $\Sigma^{\ff}$, but cannot
be extended to harmonic $(1,1)$-forms on the whole of $Y$, i.e.\ they lie in
the cokernel of $\io$.
We denote a basis of those $(1,1)$-forms by $\tilde \omega_a$,
and the full cohomology group $H^{(1,1)}(\Sigma^{\ff})$ is
split according to
\begin{equation}
H^{(1,1)}(\Sigma^{\ff})=\iota^*H^{(1,1)}(Y)\oplus\tilde H^{(1,1)}(\Sigma^{\ff})\ .
\end{equation}
As argued in \cite{Jockers:2005zy}, the basis $\tilde\omega_a$ can
always be chosen in
such a way that
\begin{equation}\label{DBRANEeq15}
\int_{\Sigma^{\ff}} \io \omega_\alpha\wedge\tilde\omega_a=0\ .
\end{equation}

%%%%%%%%%%%%%%%%%%%%%%%%%%%%%%%%%%%%%%%%%%%%%%%%%%%%%%%%%%%%%%%%%%%%%%%%%%%%%%%%

\subsection{From DBI to $D$-terms}
\label{dbidterm}

%%%%%%%%%%%%%%%%%%%%%%%%%%%%%%%%%%%%%%%%%%%%%%%%%%%%%%%%%%%%%%%%%%%%%%%%%%%%%%%%

Using a product ansatz for the spacetime, the DBI action  (\ref{DBRANEeq1})
can be rewritten as:
\begin{equation}\label{DBRANEeq7}
S_{DBI}=-\mu_7\int_{\mathcal M_4}d^4xe^{-\phi}\sqrt{-\det(g_{(4)})}
\sqrt{\det\left(1+ 2\pi\alpha' g_{(4)}^{-1} F_{(4)}\right)}\Gamma_\ff\ ,
\end{equation}
where $F_{(4)}$ denotes the four-dimensional field-strength and
$g_{(4)}$ is the (string frame) metric of $\mathcal M_4$. Further, we defined
\begin{equation}\label{DBRANEeq16}
\Gamma_\ff=\int_{\Sigma^{\ff}}d^4 z \sqrt{\det(g_{\Sigma^{\ff}}+\mathcal F)}.
\end{equation}
A low energy derivative expansion of (\ref{DBRANEeq7}) shows that the contribution
of the D7-brane to the four-dimensional scalar potential is given by
\begin{equation}
V_{D7}=\mu_7\, e^{-\phi}\Gamma_\ff (\V e^{-2\phi})^{-2}
= \mu_7\, e^{3\phi} \V^{-2} \Gamma_\ff\ . \label{Pot}
\end{equation}
The factor $(\V e^{-2\phi})^{-2}$ appears
in the potential after transforming to the four-dimensional
Einstein frame ($g_{\mu \nu}^{\rm (str)} = g_{\mu \nu}^{\rm (E)}
(\V e^{-2\phi})^{-1}$), where the (dimensionless)
volume of the Calabi-Yau orientifold, measured in the ten-dimensional
string frame metric, is given by\footnote{We
use $\sqrt{\alpha'}$ as the unit of length.}
\begin{equation}
\V = \frac16 (\sqrt{\alpha'})^{-6} \int_Y J \wedge J \wedge J = \frac16
\mathcal K_{\alpha\beta\gamma} v^\alpha v^\beta v^\gamma\ .
\end{equation}
Here $\mathcal K_{\alpha\beta\gamma}$ are the triple intersection numbers
\begin{equation} \label{tripel}
\mathcal K_{\alpha\beta\gamma}
=(\sqrt{\alpha'})^{-6}\int_Y\omega_\alpha\wedge\omega_\beta\wedge
\omega_\gamma\ .
\end{equation}
For future reference, we note that the reduced Planck mass after the
Weyl transformation to the four-dimensional Einstein frame is given by
\begin{equation} \label{mp}
M_{\rm P}^2 = \frac{\alpha^{\prime 3}}{\kappa_{10}^2} = 2 (2 \pi)^{-7} \alpha^{\prime -1}\ .
\end{equation}

In addition to (\ref{Pot}), the low energy expansion
of (\ref{DBRANEeq7}) also determines the gauge coupling to be
\begin{equation} \label{gc}
g_\ff^{-2}=\mu_7(2\pi\alpha')^2e^{-\phi}\Gamma_\ff\ .
\end{equation}
Thus both, the contribution to the potential and to the gauge coupling,
are determined by $\Gamma_\ff$. In order to proceed further,
we therefore have to calculate (\ref{DBRANEeq16}). In similar situations with D6-branes
at angles (or, in the T-dual picture,  D9-branes with world-volume fluxes), different methods
were applied in the past. In \cite{Cremades:2002te,Blumenhagen:2003vr}
each brane stack is separately chosen to keep some supersymmetry intact. This
allows to make use of (\ref{DBRANEeq3}), where a priori each brane stack might be
calibrated using a different $\theta$-phase. However, only if all branes
are calibrated with the same $\theta$ as the O-planes, i.e.\ using $\theta=0$
\cite{Blumenhagen:2003vr},
an overall supersymmetry is unbroken; otherwise, a $D$-term potential arises.

An alternative method to treat the analogue of (\ref{DBRANEeq16})
in the case of D6-branes
at angles in a toroidal orientifold
has recently been applied in \cite{Villadoro:2006ia}.
As the metric and world-volume fluxes are known
explicitly in toroidal orientifolds, the integral analogous to
(\ref{DBRANEeq16}) can be
calculated explicitly and expanded around a supersymmetric vacuum, which also allows
to read off the $D$-term potential. It turns out that this method can not only
be carried over to the D7-brane case, but
it is also possible to apply it to a general
Calabi-Yau for which the metric and world-volume fluxes are not explicitly known.
We defer that calculation to appendix \ref{alternative} and here proceed
with a variant of the method using (\ref{DBRANEeq3}) and (\ref{DBRANEeq4})
as a starting point.\footnote{Still another method to calculate the $D$-term
potential in the case of D7-branes was applied in \cite{Jockers:2005zy}. The
authors determined the $D$-term both by considering the gauging of sigma-model
symmetries after turning on world-volume fluxes, as well as by
analyzing the fermionic terms on the world-volume of the D7-branes. We here
reproduce their result directly from a reduction of the bosonic DBI action.}

Due to (\ref{DBRANEeq4}) we can expand the 2-form flux $\mathcal F$ into
basis elements of  $H^{(1,1)}(\Sigma^{\ff})$:\footnote{Note that our $\tilde f$ is
not identical to the one used in \cite{Jockers:2005zy}, because we do not
expand the world-volume flux on the brane and its mirror simultaneously. Thus,
in the end we still have to sum over the branes and their mirror images. The end result
is, however, equivalent to the one of \cite{Jockers:2005zy}.}
\begin{equation}\label{DBRANEeq6}
\mathcal F=f^\alpha\iota^*\omega_\alpha+\tilde f^a\tilde\omega_a\ .
\end{equation}
To avoid any misunderstanding, let us stress that we only consider
fluxes on the D7-brane wrapped on $\Sigma^{\ff}$ and thus $\iota^*$
always refers to the pullback onto the world-volume of this brane (we did
not want to overload the notation with an index $\ff$ on $\iota^*$ though).
In particular, $f^{\alpha} \neq 0$ denotes the $\iota^*\omega_\alpha$-component
of the flux on D7$_{\ff}$ and not a flux on another possibly present  D7-brane wrapped
on the cycle $\Sigma^\alpha$.

Now using (\ref{DBRANEeq3}) in (\ref{DBRANEeq16}), we obtain
\begin{equation}
\Gamma_\ff=\tilde \Gamma_\ff e^{-i\theta}=|\tilde\Gamma_\ff| e^{i(\tilde\theta-\theta)}
\end{equation}
with
\begin{equation} \label{gtf}
\tilde \Gamma_\ff \equiv \frac{1}{2} \int_{\Sigma^{\ff}}\left(\iota^*J\wedge\iota^*J
-\mathcal F\wedge\mathcal F\right)+i\int_{\Sigma^{\ff}}
\left(\iota^*J\wedge\mathcal F\right)\ .
\end{equation}
The phase $\theta$ is fixed since the tension of the brane should
be real and positive. This gives the relation
\begin{equation}
\theta=\tilde \theta+2 n\pi,
\end{equation}
where $n\in \mathbb Z$ and
\begin{equation}\label{DBRANEeq12}
\tilde\theta=\arctan\left(\frac{2\int_{\Sigma^{\ff}}\iota^*J\wedge\mathcal F}{\int_{\Sigma^{\ff}}
(\iota^*J\wedge\iota^*J-\mathcal F\wedge\mathcal F)}\right)
\end{equation}
is the phase of $\tilde \Gamma_\ff$. Thus, we deduce that
\begin{equation}\label{DBRANEeq40}
\Gamma_\ff=|\tilde\Gamma_\ff|=\sqrt{\left(\frac{1}{2}\int_{\Sigma^{\ff}}(\iota^*J\wedge\iota^*J
-\mathcal F\wedge\mathcal F)\right)^2+\left(\int_{\Sigma^{\ff}}\iota^*J\wedge\mathcal
F\right)^2}\ .
\end{equation}
We can use (\ref{DBRANEeq5}), (\ref{DBRANEeq15}) and (\ref{DBRANEeq6})
to express $\Re\tilde\Gamma_\ff$ and $\Im\tilde\Gamma_\ff$ as
\begin{equation}
\begin{split}
\Re\tilde\Gamma_\ff&=\frac{1}{2}\int_{\Sigma^{\ff}}(\iota^*J\wedge\iota^*J
-\mathcal F\wedge\mathcal F)=\Big( \frac{1}{2} v^\alpha v^\beta
\mathcal K_{\alpha\beta\ff}-\mathfrak{f}_\ff\Big) (\sqrt{\alpha'})^4 \ ,\\
\Im\tilde\Gamma_\ff&=
\int_{\Sigma^{\ff}}\iota^*J\wedge\mathcal F= v^\alpha Q_{\alpha \ff}\,
(2 \pi)^2 \alpha^{\prime 2}\ . \label{tildegammareim}
\end{split}
\end{equation}
Here, $\mathcal K_{\alpha\beta\ff}$ is the triple intersection number (\ref{tripel})
of the four-cycle $\Sigma^{\ff}$ and the Poincar{\'e} dual 4-cycles of
$\omega_\alpha$ and $\omega_\beta$, and we used the abbreviations
\begin{equation} \label{Q}
Q_{\alpha \ff}= \alpha^{\prime -2}
\int_{\Sigma_{\ff}}\iota^*\omega_\alpha \wedge \frac{\mathcal F}{(2\pi)^2}
= \alpha^{\prime -1}
\int_{\Sigma_{\ff}}\iota^*\omega_\alpha \wedge \frac{F}{2\pi}
= (2\pi)^{-2} f^\beta \mathcal K_{\alpha\beta\ff}\ ,
\end{equation}
and
\begin{equation} \label{fhat}
\mathfrak{f}_\ff=\frac12 \Big(f^\alpha f^\beta\mathcal K_{\alpha\beta\ff}+\tilde f^a\tilde f^b\mathcal
K^{(\ff)}_{ab}\Big)
\end{equation}
with
\begin{equation}
\mathcal K^{(\ff)}_{ab}=\alpha^{\prime -2}
\int_{\Sigma^{\ff}}\tilde\omega_a\wedge\tilde\omega_b\ .
\end{equation}
Note that the part of $\mathcal F$ that is only harmonic
on $\Sigma^{\ff}$, i.e.\ $\tilde f^a$, does not appear in $\Im\tilde\Gamma_\ff$ but
only in $\Re\tilde\Gamma_\ff$. Moreover, $Q_{\alpha \ff}$, as defined in (\ref{Q}), is
integer valued because $F$ is quantized in such a way that its integral over an arbitrary
2-cycle is a multiple of $2\pi$, i.e.\
\begin{equation}
\int F = 2 \pi n\ , \quad n \in \mathbb{Z}\ .
\end{equation}

As mentioned above, the condition for the D7-brane to preserve the same supersymmetry
as the O7-plane enforces $\theta=\tilde\theta=0$.
Using (\ref{DBRANEeq12}) this translates into the condition
\begin{equation}
\Im \tilde\Gamma_\ff=\int_{\Sigma^{\ff}}\iota^*J\wedge\mathcal F
=0\ .
\end{equation}
Allowing for small supersymmetry breaking, the expansion of (\ref{DBRANEeq12}) yields,
for small $\Im \tilde\Gamma_\ff$,
\begin{equation}
\theta\sim \Im \tilde\Gamma_\ff.
\end{equation}
%\footnote{Since $\iota^*J$ and $\mathcal F$ can be expanded into harmonic
%forms and $\Sigma$ is compact, $\int_\Sigma\iota^*J\wedge \mathcal F=0$ is
%identical to $\iota^*J\wedge \mathcal F=0$.}
Following \cite{Villadoro:2005yq}, we then expand (\ref{DBRANEeq40})
in the limit $|\Im\tilde\Gamma_\ff| \ll |\Re\tilde\Gamma_\ff|$, and the
contribution (\ref{Pot}) of the D7-brane to the potential becomes
\begin{eqnarray}
\mu_7\, e^{3\phi} \V^{-2} \Gamma_\ff&=&
\mu_7\, e^{3\phi} \V^{-2}
\Re\tilde\Gamma_\ff\sqrt{1+\left(\frac{\Im\tilde\Gamma_\ff}{\Re
\tilde\Gamma_\ff}\right)^2}\nonumber\\
&\approx&
\mu_7\,
e^{3\phi} \V^{-2} \Re\tilde\Gamma_\ff+\frac12 \mu_7\,
e^{3\phi} \V^{-2}
\frac{1}{\Re\tilde\Gamma_\ff}(\Im\tilde\Gamma_\ff)^2\ .\label{DBRANEeq17}
\end{eqnarray}
The first term proportional to $\Re {\tilde{\Gamma}}_{\ff}$
is a tension term that is cancelled if one sums over all contributions of
D7/D3-branes, O7/O3-planes, 3-form fluxes and possibly $R^2$-terms
due to RR-tadpole cancellation condition. This was discussed in similar cases for
instance in
\cite{Giddings:2001yu,Blumenhagen:2003vr,Villadoro:2005cu}. The second
term we would like to interpret as a contribution to the $D$-term potential.
To bring it into
the standard supergravity form we need an explicit formula for the D7-brane
gauge coupling
and the definition of the right field theoretic
variables, i.e.\ those in which the
gauge kinetic function becomes holomorphic and in which the sigma model
metric is manifestly  K{\"a}hler.

The gauge coupling is obtained  from (\ref{gc}) using again the  expansion
 (\ref{DBRANEeq17}) and keeping the first nontrivial term, which this time is the
first one proportional to $\Re {\tilde{\Gamma}}_{\ff}$. This leads to
\begin{equation}\label{DTERMeq5a}
g_\ff^{-2}= \mu_7(2\pi\alpha')^2 e^{-\phi}\Re\tilde\Gamma_\ff
= \mu_7(2\pi)^2 \alpha^{\prime 4} \Big( e^{-\phi} \frac{1}{2} v^\alpha v^\beta
\mathcal K_{\alpha\beta\ff}- e^{-\phi}\mathfrak{f}_\ff\Big).
\end{equation}
Comparing with (\ref{mu7}), we see that the $\alpha'$-factors drop out
and the gauge coupling is (classically) dimensionless as it should be in
four dimensions. In particular,
\begin{equation} \label{mu7b}
\mu_7(2\pi)^2 \alpha^{\prime 4} = \frac{1}{(2 \pi)^5}\ .
\end{equation}
Moreover, the first term in parenthesis in (\ref{DTERMeq5a})
can be rewritten as
\begin{equation}
e^{-\phi} \frac{1}{2} v^\alpha v^\beta \mathcal K_{\alpha\beta\ff} =
\frac{1}{2} \hat v^\alpha \hat v^\beta \mathcal K_{\alpha\beta\ff} =
\partial_{\hat v^\ff} \hat \V \equiv \hat \V^\ff\ ,
\end{equation}
where
\begin{equation} \label{CYvolume}
\hat \V = \frac16 \hat v^\alpha \hat v^\beta \hat v^\gamma
\mathcal K_{\alpha\beta\gamma} = e^{-3 \phi /2} \V
\end{equation}
is the volume of the Calabi-Yau orientifold as
measured in the ten-dimensional Einstein frame metric
(which is related
to the string frame metric by the usual dilaton factor
$g_{MN}^{\rm (E)} = g_{MN}^{\rm (str)} e^{-\phi/2}$, as appropriate for
ten dimensions).
The corresponding
2-cycle moduli are given by
\begin{equation}
\hat v^\alpha = e^{-\phi/2} v^\alpha\ .
\end{equation}
Up to normalization, $\hat \V^\ff$ is the real part of the    K{\"a}hler modulus
$T^{\ff}$ of the wrapped  4-cycle $\Sigma^{\ff}$,
and $e^{-\phi}$ is the real part of the dilaton, cf.\ for instance
\cite{Becker:2002nn}.\footnote{We differ from the definition in
\cite{Becker:2002nn} by factors of $i$ and the arbitrary
normalization.}
Thus, choosing the normalization appropriately, (\ref{DTERMeq5a}) can be rewritten as
\begin{equation} \label{realpart}
g_\ff^{-2} =  \Re T^\ff -
\mathfrak{f}_\ff \Re S\ .
\end{equation}
The imaginary parts of the moduli $T^\ff$ and $S$ are given by
RR-fields as can be seen from the
reduction of the CS terms on the D7-brane
(more precisely, they are given by
$C_0$ for $S$ and $a^\ff=\alpha^{\prime -2}
\int_{\Sigma^{\ff}}\iota^*C^{(4)}$
for $T^\ff$). The relative signs are determined by demanding
a holomorphic gauge kinetic function (cf.\ appendix \ref{gaugekin} for more details), and we arrive
at\footnote{The same relative signs were found for instance
in \cite{Andrianopoli:2003jf}, cf.\ their formula (8).
Note that their variables differ by factors of $\pm i$ from ours.}
\begin{equation} \label{kaehlermodsa}
T^\ff = \frac{1}{(2 \pi)^5} \left(\hat \V^\ff + i a^\ff\
\right) \quad , \quad
S =  \frac{1}{(2 \pi)^5} \left(e^{-\phi} - i C_0\right)\ .
\end{equation}
For general
4-cycles labelled by  $\alpha$, one obtains the analogous
 K{\"a}hler moduli
 \begin{equation} \label{kaehlermods}
T^\alpha = \frac{1}{(2 \pi)^5} \left(\hat \V^\alpha + i a^\alpha\right) \ .
%S = e^{-\phi} - i C_0\ .
\end{equation}
With these moduli one verifies that
\begin{equation} \label{vhat}
\hat v^\alpha = - \frac{2}{(2 \pi)^5} \hat \V (\partial_{T^\alpha} K)\ ,
\end{equation}
where
\begin{equation}
K = - 2 \ln \hat \V
\end{equation}
is the K{\"a}hler potential for the moduli space of the K{\"a}hler moduli, and
$\hat \V$ was defined in (\ref{CYvolume}).

Before continuing, we should note that we neglected the open string moduli
related to D7-brane fluctuations and Wilson-lines, which would otherwise also
appear in the definitions of $T^\alpha$ and $S$.
Taking them into account properly would require to consider open string
1-loop corrections in order to get holomorphic gauge kinetic functions
\cite{Berg:2004ek}.\footnote{A closed string dual version of calculating these
corrections to the gauge couplings was discussed in \cite{Giddings:2005ff,Baumann:2006th}.}
We thus assume that these moduli can be fixed, e.g.\ by the presence of fluxes
\cite{Cascales:2004qp,Camara:2004jj,Lust:2005bd}.

Now we have all the ingredients to rewrite the potential arising from
(\ref{DBRANEeq17}) in a more familiar way. As we already mentioned, tadpole cancellation
implies that the first term proportional to $\Re \tilde \Gamma$ drops out of the potential.
The remaining contribution of a single D7-brane to the $D$-term
potential is given by
\begin{eqnarray} \label{dD7}
\frac12 \mu_7\, e^{3\phi} \V^{-2}
\frac{1}{\Re\tilde\Gamma_{\Sigma^{\ff}}}(\Im\tilde\Gamma_{\Sigma^{\ff}})^2\
&=& \frac{1}{2g^{-2}_{\ff}} \Big[\frac{M_{\rm P}^2}{4 \pi^2} Q_{\alpha \ff}
(\partial_{T^\alpha}K) \Big]^2 \nonumber \\
&=&
\frac{1}{2(\Re T^\ff -
\mathfrak{f}_\ff \Re S)} \Big[ \frac{M_{\rm P}^2}{4 \pi^2} Q_{\alpha \ff}
(\partial_{T^\alpha}K) \Big]^2\ .
\end{eqnarray}
In the first equality of (\ref{dD7}) we used (\ref{mu7}), (\ref{mp}), (\ref{tildegammareim}),
(\ref{DTERMeq5a}), (\ref{CYvolume}) and (\ref{vhat}).
In the last equality we wrote out explicitly the $U(1)_{\ff}$-gauge coupling
to emphasize that it depends both on the K{\"a}hler modulus $T^{\ff}$ and the dilaton $S$. 
The dilaton dependence is often ignored in the literature. Note that using the
conventions of \cite{Polchinski:1998rr}, there is no additional contribution to (\ref{dD7})
from the orientifold image of the 
D7-brane.\footnote{There are other conventions in the literature, where
the tension and the charge is ``democratically" distributed over the branes and images.
In that case the contribution of a single D7-brane to the $D$-term potential would be half
of (\ref{dD7}), the other half would come from the image.}

Our derivation of the $D$-term made use of the Abelian DBI action (\ref{DBRANEeq1}).
If one had a stack of D7$_{\ff}$-branes instead, which does not lie on top of the
O7-planes, the gauge group would be $U(N_{\ff})$. In that case, if the world volume flux
lies in the diagonal $U(1)$-subgroup,
our calculation should carry over and the generalization of (\ref{dD7}) gets an
additional factor of $N_{\ff}$. As the same is true for the $U(1)_{\ff}$-gauge couplings
(\ref{DTERMeq5a}), the result becomes
\begin{equation} \label{DBRANEeq30}
V_D= \frac{1}{2g^{-2}_{\ff}} 
\Big[\frac{N_{\ff} M_{\rm P}^2}{4 \pi^2} Q_{\alpha \ff}
(\partial_{T^\alpha}K) +\ldots  \Big]^2\ ,
\end{equation}
%
%
%\begin{equation} \label{DBRANEeq30}
%V_D=\sum_i \frac12 \mu_7\, e^{3\phi} \V^{-2}
%\frac{2 N_i}{\Re\tilde\Gamma_{\Sigma^i}}(\Im\tilde\Gamma_{\Sigma^i})^2\
%= \sum_{i} \frac{1}{2g^{-2}_{\Sigma^i}} \Big[ - \frac{N_i M_{\rm P}^2}{4 \pi^2} Q_{\alpha {\Sigma^i}}
%(\partial_{T^\alpha}K) \Big]^2\ ,
%\end{equation}
%
where the dots stand for the additional matter contributions, 
which cannot be derived from the
DBI action, cf.~eq.~(\ref{Dterm}).
Now (\ref{DBRANEeq30}) is of the familiar form of a $D$-term potential arising
from a  gauged $U(1)$-symmetry. When compared with the expressions
(\ref{TTq}), (\ref{Lag}) and (\ref{Dterm}), 
the coefficients $\frac{N_{\ff}}{4 \pi^2} Q_{\alpha\ff}$
correspond, for $\alpha=\cc$, to the constant Killing vector $-i\eta^{T^\cc}=q$
of the gauged shift symmetry, at least up to a sign which can not be determined from
(\ref{DBRANEeq30}) alone as it is quadratic in $q$.

%%%%%%%%%%%%%%%%%%%%%%%%%%%%%%%%%%%%%%%%%%%%%%%%%%%%%%%%%%%%%%%%%%%%%%%%%%%%%%%%

\subsection{$U(1)$ from CS}

%%%%%%%%%%%%%%%%%%%%%%%%%%%%%%%%%%%%%%%%%%%%%%%%%%%%%%%%%%%%%%%%%%%%%%%%%%%%%%%%

The presence of the gauged $U(1)$ symmetry can also be inferred
by looking at the part
\begin{equation}
-\frac{\mu_7}{2}\int_{\mathcal W}\iota^*C_4\wedge\mathcal F\wedge \mathcal F,
\end{equation}
of the CS action. Taking one of the $\mathcal F$ in this expression to be along $\Sigma^{\ff}$ (i.e.\
it is part of the world volume flux) and one as the field strength of the four-dimensional
$U(1)$-field on the D7-brane stack, and expanding
\begin{equation}
C_4=C_2^\alpha \wedge \omega_{\alpha} + \ldots,
\end{equation}
gives
\begin{equation} \label{bfterm}
-\mu_7\int_{\Sigma^{\ff}}  \iota^*\omega_\alpha\wedge\mathcal F \int_{\mathcal M_4}
C_2^{\alpha}\wedge F=- \mu_7 (2 \pi)^2 \alpha^{\prime 2} Q_{\alpha \ff} \int_{\mathcal M_4}
C_2^{\alpha}\wedge F\ .
\end{equation}
Hence, if $Q_{\alpha \ff}\neq 0$ the derivative
of the axion $a^\alpha$ dual to
the 2-forms $C_2^\alpha$ will be covariantized, just as in (\ref{Lag}).
{}From the definition of $Q_{\alpha \ff}$ in (\ref{Q}), it is obvious that this
happens if the 4-cycle $\Sigma^\alpha$ (which is dual to $\omega_\alpha$) and $\Sigma^{\ff}$
intersect over a 2-cycle on which the world-volume flux is non-trivial.
In particular, when  $\Sigma^{\ff}$ has self-intersections,
$T^\ff$ can get charged in this way, which is the case discussed in
\cite{Burgess:2003ic,Jockers:2005zy}.

If the above requirement is fulfilled and a charged K{\"a}hler modulus $T^{\alpha}$ exists,
the corresponding gauged  $U(1)_{\ff}$
will become  anomalous if  the 4-cycle $\Sigma^\alpha$ dual to $\omega_{\alpha}$
is wrapped by a D7-brane (or a brane stack).
More precisely, the tree level effective action contains the term
 \begin{equation}\label{inv}
\int_{\mathcal{M}_{4}} \Im(T^\alpha) \tr F^\alpha \wedge F^\alpha
 \end{equation}
with $F^{\alpha}$ being the field strength on the branes 
wrapped around the 4-cycle dual to $\omega^{\alpha}$ (we denote the gauge group 
of these branes by $G_\alpha$). This seems to break the 
invariance under local shifts of $\Im(T^\alpha)$, i.e., the gauge symmetry $U(1)_{\ff}$.
At the quantum level, however, there are also mixed triangle anomalies
of the type $U(1)_{\ff} - [G_{\alpha}]^{2}$  due to
chiral bifundamental fermions from open strings stretching between the two  
intersecting branes
(whose existence also depends on the same condition $Q_{\alpha \ff}\neq 0$, as we
elaborate on in the next section). These triangle anomalies
cancel the $U(1)_{\ff}$ non-invariance of (\ref{inv})
by a Green-Schwarz mechanism.
It is precisely these anomalous bifundamental matter fields\
that also modify
the naive gaugino condensation superpotential to yield the ADS superpotential (\ref{wads}),
which is then naturally invariant under $U(1)_{\ff}$ transformations,  
as required for the mechanism suggested in
\cite{Achucarro:2006zf}.
A more detailed analysis of the gauge invariance of the superpotential
is subject of section \ref{anomalies}.

%%%%%%%%%%%%%%%%%%%%%%%%%%%%%%%%%%%%%%%%%%%%%%%%%%%%%%%%%%%%%%%%%%%%%%%%%%%%%%%%

\subsection{Curvature corrections}
\label{R2}

%%%%%%%%%%%%%%%%%%%%%%%%%%%%%%%%%%%%%%%%%%%%%%%%%%%%%%%%%%%%%%%%%%%%%%%%%%%%%%%%

Before moving on to that discussion, let us come back to the issue of curvature
corrections.
%to the
%DBI (and CS) action, (\ref{DBRANEeq1}) and (\ref{DBRANEeq2}), and
%the calibration conditions (\ref{DBRANEeq3}).
We focus here on the calibration condition (\ref{DBRANEeq3}).
This implies
% %
% \begin{equation}
% \cos \theta\, J \wedge \cf + \sin \theta \left( \frac12 \cf\wedge \cf
% - \frac12 J \wedge J \right) = 0\ ,
% \end{equation}
% %
% which itself can be re-expressed as
% %
\begin{equation}
\Im \left(e^{-i \theta} \Phi \right) = 0\ ,
\end{equation}
with
\begin{equation}
\Phi = \frac12 (\cf-i \io J) \wedge (\cf-i \io J) \ .
\end{equation}
For the sake of the generalization to include the higher curvature terms,
the following form is more convenient:
\begin{equation}
\Im \left. \left(e^{-i \theta} e^{\cf - i \io J} \right)\right|_{\rm top} = 0\ ,
\end{equation}
where ``top'' denotes the projection of the form to its top degree on the
4-cycle, i.e.\ to the 4-form part.

In the analogous case of $D9$-branes with fluxes the inclusion of the higher curvature
terms (in the large volume limit) has been argued to be \cite{Douglas:2000ah,Blumenhagen:2005zh}
\begin{equation}
\Im \left. \left(e^{-i \theta} e^{\cf - i J} \sqrt{\hat{A}(T(Y))}
\right)\right|_{\rm top} = 0\ ,
\end{equation}
where $\hat{A}(T(Y))$ is the A-roof genus \cite{Eguchi:1980jx}
\begin{equation} \label{aroof}
\hat{A}(T(Y)) = 1 - \frac{1}{24} p_1(T(Y)) + \ldots \ ,
\end{equation}
with $p_1$ the first Pontryagin class and $T(Y)$ the tangent bundle
of the whole internal manifold.
In the case at hand, i.e.\ for D7-branes wrapped around a 4-cycle
$\Sigma$,
it is plausible to conjecture the corresponding generalization
\begin{equation} \label{generalized}
\Im \left. \left(e^{-i \theta} e^{\cf - i\io J}
\sqrt{\frac{\hat{A}(T(\Sigma))}{\hat{A}(N(\Sigma))}}
\right)\right|_{\rm top} = 0\ ,
\end{equation}
where $T(\Sigma)$ and $N(\Sigma)$ are the tangent- and normal bundle of $\Sigma$.
We will see further evidence for this conjecture in a moment.
It is straightforward to expand the bracket, leading to
\begin{equation}
\Im \left(e^{-i \theta} \left(\frac12 (\io J + i \cf ) \wedge (\io J + i \cf ) + \frac{1}{48}
p_1(T(\Sigma)) - \frac{1}{48} p_1(N(\Sigma)) \right) \right) = 0\ ,
\end{equation}
where we multiplied the left hand side of (\ref{generalized})
by minus one in order to facilitate comparison with earlier formulas.
We note that the inclusion of the higher derivative corrections amounts
to a shift in $\tilde \Gamma_{\ff}$ of (\ref{gtf}) according to
\begin{equation} \label{tildeGF}
\tilde \Gamma_{\ff} \rightarrow \tilde \Gamma_{\ff} + \frac{1}{48}
\int_{\Sigma^{\ff}} \Big( p_1(T(\Sigma^{\ff})) - p_1(N(\Sigma^{\ff}))\Big)\ .
\end{equation}
As the additional contribution is real, the imaginary part of 
$\tilde \Gamma_{\ff}$ does not change. This does, however, not mean that 
the D-term potential is not changed either. In fact it is known that the 
K\"ahler potential does get corrections
from higher curvature terms, at least from those arising in
the bulk action \cite{Becker:2002nn}. These amount to a shift in the argument 
of the K\"ahler potential for the K\"ahler moduli, i.e.
\begin{equation} \label{correctedK}
K = -2 \ln \Big(\hat \V + \hat \chi e^{-3 \phi /2}\Big)\ ,
\end{equation}
where $\hat \chi$ is a constant proportional to the Euler number of the compactification 
space. Using this in (\ref{vhat}), one would obtain instead
\begin{equation}
\hat v^\alpha = - \frac{2}{(2 \pi)^5} \Big(\hat \V+ \hat \chi e^{-3 \phi /2}\Big) 
(\partial_{T^\alpha} K)\ .
\end{equation}
However, also the Einstein-Hilbert term in string frame is modified by the higher derivative 
corrections, leading to a term proportional to 
\begin{equation}
e^{-2\phi} (\V + \hat \chi) R\ .
\end{equation}
Thus, going to the four-dimensional Einstein frame now amounts to 
rescaling the four-dimensional metric by $e^{-2 \phi} (\V + \hat \chi)$, so that 
the left hand side of (\ref{dD7}) would contain a factor of
$(\hat \V + \hat \chi e^{-3 \phi/2})^{-2}$ now instead of $e^{3\phi} \V^{-2}= \hat \V^{-2}$. 
Thus, the right hand side of 
(\ref{dD7}) still holds, even if the K\"ahler potential is modified by the 
higher derivative corrections as in (\ref{correctedK}). In particular the charges of 
the axions are unchanged.\footnote{We should note that implicit in our argument is 
the assumption that the D-term and also the gauge couplings are still determined
by the real and imaginary parts of $\tilde \Gamma_{\ff}$. It would be nice 
to have a conclusive derivation of this by including the higher derivative 
corrections on the brane also on the right hand side of (\ref{DBRANEeq3}).}  

One can give an additional argument that these charges 
are not modified by repeating the analysis of the last subsection, including the
known curvature terms in the CS-action.
The corrected CS-action is given by \cite{Green:1996dd,Cheung:1997az,Minasian:1997mm}:
\begin{equation} \label{CScorr}
S_{\rm CS}=-\mu_7\int_{\mathcal W}\sum_p \iota^*C_p\wedge e^{\mathcal F}\wedge
\sqrt{\frac{\hat A(T(\Sigma^{\ff}))}{\hat A(N(\Sigma^{\ff}))}}\ ,
\end{equation}
where, as usual, the integral picks out the 8-form part of
the integrand. As the axions
charged under $U(1)_{\ff}$ originate as the duals of $C_2^\alpha$, one is
interested in those terms of (\ref{CScorr}) that contain $C_4$. This, however,
does not allow the curvature corrections to modify the charge of the axions, as
the expansion of $\hat{A}$ contains only forms of degree $4 n$
(for $n \in \mathbb{N}$) and one also needs one power of the gauge field strength along the
non-compact space-time.

In contrast to the $D$-term, the tension and gauge coupling
are related to the real part of $\tilde \Gamma_{\ff}$ and,
therefore, do receive corrections from (\ref{tildeGF}). 
If the tension is modified,
also the tadpole condition gets a contribution from
the higher curvature terms. This is the case if there are 
D7-branes wrapped around
cycles $\Sigma$ for which the integrals of
$p_1(T(\Sigma))$ and/or $p_1(N(\Sigma))$ over $\Sigma$ are 
non-vanishing, cf.\ a
related discussion in \cite{Giddings:2001yu}.

We see that the additional contribution to the real part of 
$\tilde \Gamma_{\ff}$
can be taken into account by adjusting the definition (\ref{fhat}) to
\begin{equation} \label{hatfrakf}
\hat{\mathfrak{f}}_{\ff}=\mathfrak{f}_{\ff}
-\frac{\alpha^{\prime -2}}{48}\int_{\Sigma^{\ff}}\Big(p_1(T(\Sigma^{\ff}))
-p_1(N(\Sigma^{\ff}))\Big)\ ,
\end{equation}
which leads to a modification of the
gauge kinetic function to
\begin{equation}
f^{\ff} = T^{\ff} - \hat{\mathfrak{f}}_{\ff} S\ . \label{fff}
\end{equation}
In appendix \ref{gaugekin}, we verify that this is consistent with
a reduction of the known higher curvature corrections to the
CS-part of the D7-brane action (which gives an independent
derivation of the corrections to the imaginary part of $f^{\ff}$).
This supports the proposal (\ref{generalized}) and the method  
of including the higher derivative corrections on the brane by
modifying $\tilde \Gamma_{\ff}$, according to (\ref{tildeGF}), 
in the derivation of section \ref{dbidterm}.

%%%%%%%%%%%%%%%%%%%%%%%%%%%%%%%%%%%%%%%%%%%%%%%%%%%%%%%%%%%%%%%%%%%%%%%%%%%%

\section{Anomalies and gaugino condensation on D7-branes}
\label{anomalies}

%%%%%%%%%%%%%%%%%%%%%%%%%%%%%%%%%%%%%%%%%%%%%%%%%%%%%%%%%%%%%%%%%%%%%%%%%%%

Whereas the previous section primarily focused on the generation of the $D$-terms
due to the world volume fluxes on the D7-branes, the present section is devoted to the
second important ingredient of the construction of \cite{Achucarro:2006zf},
namely the gaugino condensation. The connecting link between these two
sub-effects is provided by certain types of anomalies. We therefore start with an
analysis of the relevant anomalies that might occur in our set-up.
To this end, let us  consider, just  as  in the introduction,
  two different stacks of D7-branes, one denoted
by D7$_\cc$ and  the other one by D7$_\ff$.
On stack D7$_\cc$,  we assume a gauge group
$G_{\cc}$ that can undergo gaugino condensation.
As generalizations of the ADS superpotential
in the presence of (anti-)symmetric tensor representations
  are not very well
understood,\footnote{See, however, \cite{Csaki:1996zb}.}   we assume that  
these representations are absent. The simplest situation in which this is the case is when
 the D7$_{\cc}$ stack does not intersect the O-planes, which will  henceforth
 be assumed. The gauge group   $G_{\cc}$  will therefore be
  of the form $U(N_{\cc})\cong SU(N_\cc)\times U(1)_{\cc}$.

The other stack, D7$_{\ff}$,  has a gauge group
 $G_{\ff}$ that  we require to contain  at least one  $U(1)$ factor
denoted by $U(1)_{\ff}$. 
In the generic case, when  the D7$_{\ff}$ stack does not lie on top of the 
O7-planes, one has the usual unitary gauge group including an Abelian $U(1)_\ff$ 
factor.  On the other hand,   in the  case when  the D7$_{\ff}$ stack coincides   
with  an O7-plane,  $G_{\ff}$ becomes enhanced to a symplectic or orthogonal 
gauge group. This group   can be broken to a unitary group with Abelian 
factors by switching on appropriate
  world volume fluxes, so this might a priori also be a valid option.  
However, we want to ensure at the same time that 
tensor representations of $SU(N_{\cc})$ are absent and that the 
K\"ahler modulus $T^{\cc}$ is charged under $U(1)_{\ff}$.
This is only guaranteed  when the D7$_{\ff}$ branes are \emph{not} on top of 
O7-planes, which is therefore the case we assume in the rest of this paper.    
The reason for this is  that
 the D7$_{\cc}$-stack has to intersect the D7$_{\ff}$ stack in order 
for the K{\"a}hler modulus $T^{\cc}$ to become charged under $U(1)_{\ff}$ 
so as to generate a $T_{\cc}$-dependent $D$-term. But when the D7$_{\ff}$ 
stack is on top of an O7-plane, the D7$_\cc$ stack would then  also intersect 
the O7-plane, contrary to our earlier assumption regarding D7$_{\cc}$.

\begin{figure}[h]
\begin{center}
\psfrag{A}[r][bl]{$SU(N_\cc)$}
\psfrag{B}[r][bl]{$SU(N_\cc)$}
\psfrag{C}[l][bl]{$U(1)_{\ff}$}
\includegraphics[width=0.3\textwidth]{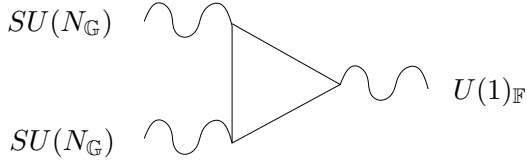}
%\qquad \qquad \quad \qquad \qquad
%\psfrag{A}[r][bl]{$U(1)_\cc$}
%\psfrag{B}[r][bl]{$U(1)_\cc$}
%\psfrag{C}[l][bl]{$U(1)_\ff$}
%\includegraphics[width=0.3\textwidth]{triangle.eps}
\caption{Mixed triangle graph.}
\label{Triangle}
\end{center}
\end{figure}

In this section, we   are interested in the compatibility
of the flux-induced  $D$-term potential
  and
the non-perturbative superpotential involving the K{\"a}hler modulus $T^\cc$
in the case where  this modulus is charged under $U(1)_{\ff}$. 
As explained in the Introduction, this compatibility is equivalent
to the cancellation of the mixed anomaly of the type
 $U(1)_{\ff}-[SU(N_{\cc})]^{2}$ by the Green-Schwarz 
mechanism.\footnote{Note that in our D-brane setup this would imply 
the cancellation of the mixed anomaly
 $U(1)_{\ff}-U(1)_{\cc}^{2}$ at the same time.}
%  As we also have a $U(1)_{\cc}$ factor in
%$G_{\cc}$, there is an analogous mixed  anomaly of the type $U(1)_{\ff}-[U(1)_{\cc}]^{2}$, which we   will   discuss along with the non-Abelian one.
The  triangle diagram of this  anomaly is
 depicted in figure
\ref{Triangle}.
%We use  (N) for the non-Abelian    $U(1)_{\ff}-[SU(N_{\cc})]^{2}$ anomaly  and  (A) for the Abelian diagram corresponding to
%$U(1)_{\ff}-[U(1)_{\cc}]^{2}$.
Other types of anomalies will in general also
appear and have to be cancelled in a complete global model, but as those other
 anomalies are not relevant for our discussion of gaugino condensation and $D$-terms, we will not consider them in this paper.

%In the remainder, we concentrate
%on the above mixed anomalies, as these are the only  anomalies that are relevant
%for our discussion.
%A generalization to more
%complicated cases (e.g.\ if $T^{\cc}$ is charged under several anomalous $U(1)$'s)
%should be straightforward.

The fields running in the loop are the bifundamental fermions
that arise at the intersections of stack D7$_{\ff}$ and D7$_{\cc}$.
If D7$_{\cc}$ intersects other branes, there might be additional
fields transforming in the (anti-)fundamental representation of $SU(N_{\cc})$.
Denoting by $N_{F}$ the total number of flavors of $SU(N_{\cc})$,
a non-perturbative superpotential is generated
by  either   instantons or gaugino condensation if 
$N_F < N_\cc$.\footnote{See, however, the example in eq.\ (4.43) and (4.44) 
of \cite{Kaplunovsky:1994fg}.}

More precisely, after integrating out the non-Abelian
gauge fields of $SU(N_\cc)$
(i.e.\ below the gaugino condensation scale), the
term in the effective Lagrangian proportional to
$\tr(\langle F_{\mu \nu} \tilde F^{\mu \nu} \rangle)$ is
\begin{equation} \label{fftilde}
\Big( (N_\cc - N_F) {\rm Arg}(\langle \lambda^a \lambda^a \rangle)
+ {\rm Arg}(\det M) + 8 \pi^2 \Im(T^\cc) \Big) \tr(\langle F_{\mu \nu} \tilde F^{\mu \nu}
\rangle)\ .
\end{equation}
%
%Actually, the same terms also appear in the case of (A)=(N)=0, the difference being
%the way in which the second and third term in brackets transform under $U(1)_{\ff}$.
The presence of the terms in (\ref{fftilde}) follows from the Veneziano-Yankielowicz-Taylor
superpotential \cite{Veneziano:1982ah,TVY,Taylor:1985fz}, i.e.\ they appear in
\begin{equation}
\int d^2 \theta\, W_{\rm VYT} + c.c.
\end{equation}
with
\begin{equation} \label{WVYT}
W_{VYT} = \frac{U}{4} f^{\cc} 
+ \frac{U}{32 \pi^2} \Big( (N_\cc - N_F) \ln \frac{U}{\Lambda^3}
+ \ln \frac{\det M}{\Lambda^{2 N_F}} \Big)\ ,
\end{equation}
where $\Lambda$ is the UV cutoff scale, $U$ is the gaugino bilinear field and
we generalized (\ref{fftilde}) slightly by allowing for a general gauge kinetic
function $f^{\cc}$.\footnote{For an expansion of $U$ in components, see
for example \cite{Burgess:1995kp}.} If one now integrates out
$U$ by its equation of motion, as
for instance in \cite{Amati:1988ft,Kaplunovsky:1994fg,Burgess:1995kp}, one
ends up with the ADS superpotential (\ref{wads}), that we repeat here for convenience,
\begin{equation} \label{wadsagain}
 W_{ADS} = \left( \frac{e^{- 8 \pi^2 f^{\cc}}}{\det M} \right)^{\frac{1}{N_\cc-N_F}}
= A(z) \left( \frac{e^{- 8 \pi^2 T^{\cc}}}{\det M} \right)^{\frac{1}{N_\cc-N_F}}\ .
\end{equation}
The second term in (\ref{fftilde}), involving the meson field matrix
\begin{equation} \label{mesons}
M_i^j = \tilde \Phi^{ja} \Phi_{ia}\ ,
\end{equation}
is necessary to ensure an anomaly free $U(1)_R$ symmetry ($\langle \lambda^a \lambda^a \rangle$
has R-charge $2$, while $\det M$ has R-charge
$2(N_F-N_\cc)$).\footnote{For a review of SYM, see for
example \cite{Intriligator:1995au,Terning:2003th}.}
Here $a$ denotes the color index and $i,j$ the flavor index.
The last term in (\ref{fftilde}) does not transform under $U(1)_R$.

Let us now discuss the transformation of the different terms under $U(1)_{\ff}$.
The gaugini are not charged under the anomalous $U(1)_{\ff}$ and, therefore, do not transform.
The last term does transform and, therefore, also the second one has to do so,
in order to cancel the transformation of $\Im(T^\cc)$.
This is in accord with the idea that the term involving the
meson fields mimics the contribution from the triangle graph 
in figure 1  below the confinement scale,
i.e.\ after integrating out the gauge degrees of freedom. 
Given that above the confinement scale
the contribution of   figure 1  cancelled the transformation of $T^\cc$ via the GS mechanism,
it is obvious that the transformation of ${\rm Arg}(\det M)$ has to perform the
same job in the effective theory below the confinement scale. Indeed, it is
clear from the definition (\ref{mesons}) that the
transformation of ${\rm Arg}(\det M)$ is proportional to the sum of the
$U(1)_{\ff}$
charges of all bifundamentals charged with respect to $SU(N_\cc)$. 
If this sum vanishes, $\det M$ does not transform but at the same time 
the diagram in figure 1 vanishes.

%This picture is also
%in agreement
%with the appearance of the terms (\ref{fftilde}) in the case (A)=(N)=0.
%In that case neither
%the second nor the third term transforms under $U(1)_{\ff}$.

With our derivation of the $D$-term potential in section \ref{section:dterm}, 
we can verify the
gauge invariance of (\ref{fftilde}) under the anomalous $U(1)_\ff$ more directly. From
(\ref{DBRANEeq30}) we read off the charge of $\Im T^\cc$ (up to a sign) to be
\begin{equation} \label{Tcharge}
\frac{N_\ff}{4 \pi^2} Q_{\cc \ff}\ ,
\end{equation}
where $N_\ff$ is the number of branes in the $\ff$-stack (not including the mirror images)
and $Q_{\cc \ff}$ is defined in
(\ref{Q}) with $\omega_\cc$ the 2-form that is Poincar{\'e} dual to the 4-cycle 
wrapped by the
$\cc$-stack.
This has to be compared to the number of bifundamental fields coming from
strings stretched between the $\ff$- and $\cc$-stacks. They each transform 
under $(SU(N_\ff),SU(N_\cc))_{U(1)_\ff}$ in the representation
\begin{equation} \label{representation}
(N_\ff, N_\cc \oplus \bar N_\cc)_{q_\ff=1}\ ,
\end{equation}
where the subscript denotes their charge under the anomalous $U(1)_\ff$ 
(we assume that the positively charged bifundamentals are the left-handed ones
appearing in the meson field (\ref{mesons}); otherwise one has 
to take their antiparticles, which would amount to a change of sign of the 
$U(1)_\ff$ charge). The fields
in the $(N_\ff, \bar N_\cc)$-representation originate from  
strings stretched from D7$_{\cc}$ to
D7$_{\ff}$, whereas the ones transforming in the $(N_\ff, N_\cc)$-representation
arise from strings stretched from the orientifold image of D7$_{\cc}$ to
D7$_{\ff}$. Obviously, with only this particle content $SU(N_{\ff})$ would be 
anomalous. Thus, additional fields charged under $SU(N_{\ff})$, for instance from
other  brane stacks that intersect D7$_{\ff}$, have to be present in a globally 
consistent model. 
As in the case of
D9-branes with fluxes, we expect that also in the case of D7-branes with fluxes the number of
chiral bifundamentals is given by the index of the Dirac operator on the intersection
of the two D7-branes and in the background of the world volume flux along this intersection.
We assume that there is no world-volume flux on the D7$_{\cc}$-stack, 
otherwise the difference
of the fluxes on the two stacks along their intersection locus
would enter the index. This would lead to different numbers of
$(N_\ff, \bar N_\cc)$- and $(N_\ff, N_\cc)$-representations, which we want to avoid.
Under this assumption, the number of $(N_\ff, \bar N_\cc)$-representations
is given by the absolute value of \cite{Eguchi:1980jx}
\begin{equation} \label{index}
{\rm index}(\nabla) = \alpha^{\prime -1} \int_{\Sigma^\ff \cap \Sigma^\cc}
\hat A (T(\Sigma^\ff \cap \Sigma^\cc)) \wedge {\rm ch} (F) = \alpha^{\prime -1}
\int_{\Sigma^\ff \cap \Sigma^\cc} \frac{F}{2 \pi} \ ,
\end{equation}
where we introduced the factors $\alpha^{\prime -1}$ for dimensional reasons,
$\hat A$ is the A-roof genus that we already encountered in
(\ref{aroof}) and in the last equality we made use of the
fact that the intersection locus $\Sigma^\ff \cap \Sigma^\cc$ is (real) two dimensional
and $\hat A$ has an expansion in forms of degree $4n$ with $n \in \mathbb{N}$. Thus
it can effectively be replaced by $1$ in our case.
We see that (\ref{index}) exactly coincides with $Q_{\cc \ff}$, cf.\ (\ref{Q}).
The number of $(N_\ff, N_\cc)$-representations is given by (\ref{index}) as well. This can be
argued as follows. As we said above, the fields transforming as $(N_\ff, N_\cc)$ come from
strings stretched from the images of D7$_{\cc}$ to D7$_{\ff}$. Thus, we have to replace
$\Sigma^\ff \cap \Sigma^\cc$ in (\ref{index}) with $\Sigma^\ff \cap \sigma(\Sigma^\cc)$,
where $\sigma(\Sigma^\cc)$ is the image of the 4-cycle $\Sigma^\cc$ under
the involution $\sigma$. However,
\begin{equation}
\int_{\Sigma^\ff \cap \sigma(\Sigma^\cc)} \frac{F}{2 \pi}
= \int_{\Sigma^\ff} \io \sigma^* \omega^\cc \wedge \frac{F}{2 \pi} =
\int_{\Sigma^\ff} \io \omega^\cc \wedge \frac{F}{2 \pi} =
\int_{\Sigma^\ff \cap \Sigma^\cc} \frac{F}{2 \pi}\ ,
\end{equation}
where we used that the Poincar{\'e}-dual of $\sigma(\Sigma^\cc)$ is given by the
$\sigma$-image (i.e.\ pullback) of $\omega^{\cc}$ (denoted $\sigma^* \omega^{\cc}$)
and in the second equality we used that $\sigma^* \omega^\cc$ and $\omega^{\cc}$ have to represent
the same cohomology class. Otherwise $(\sigma^* \omega^\cc-\omega^{\cc})$ would be
a non-trivial element of $H^2_-$ that we assumed to be vanishing.
Thus, we learn that the numbers of fundamentals and antifundamentals
of $SU(N_\cc)$ that are charged under the anomalous $U(1)_\ff$ with charge $+1$ are both given by\footnote{This equation assumes no other $SU(N_{\cc})$ matter
from possible   additional  brane stacks intersecting D7$_{\cc}$.}
\begin{equation}
N_{F}=|Q_{\cc \ff}| N_\ff.
\end{equation}
Consequently, the meson determinant transforms according to
\begin{equation}
\det M \rightarrow e^{i \epsilon 2 N_\ff |Q_{\cc \ff}| } \det M\ ,
\end{equation}
where $\epsilon$ is the gauge parameter and the factor $2$ comes from the fact that
both $\Phi$ and $\tilde \Phi$ have charge $1$ under $U(1)_\ff$. 
Thus, the charge of $\Im T^\cc$ has the right value to cancel the 
transformation of the meson determinant.

Let us finally mention that the case of no bifundamentals is the one
originally discussed in \cite{Kachru:2003aw}. In that case, as well as in the case 
that there are only non-chiral bifundamentals, the $D$-term potential would 
be independend of $T^\cc$ and, in general, a different uplift mechanism 
has to be envisaged.

\section{Conclusions}

In this paper, we studied the general compatibility of $D$-terms
from  D7-brane world volume fluxes with
gaugino condensation on D7-branes. The mutual compatibility of these
two features is crucial for the consistency of the proposal
of \cite{Burgess:2003ic} for obtaining (meta-)stable dS vacua in type IIB string theory via a
 variant of the KKLT construction \cite{Kachru:2003aw} that does not rely on the introduction
of $\overline{D3}$-branes, which break supersymmetry explicitly.

We find that in the presence of world-volume fluxes on a D7-brane (called D7$_{\ff}$ before),
wrapped around a 4-cycle $\Sigma^{\ff}$, any K{\"a}hler modulus $T^\alpha$ is charged if the following
condition is fulfilled: The 4-cycle $\Sigma^\alpha$, whose volume is measured by $T^\alpha$,
has to intersect with $\Sigma^{\ff}$ over a 2-cycle that is threaded by non-trivial
world-volume flux (on D7$_{\ff}$).\footnote{With non-trivial we mean that its integral over the
intersection 2-cycle $\Sigma^{\ff} \cap \Sigma^\alpha$ does not vanish.}
If the cycle $\Sigma^\alpha$ is wrapped by
a (stack of) D7-brane(s) as well (denoted by D7$_\alpha$), this is exactly the same condition that
ensures the presence of chiral matter from strings stretching between D7$_{\ff}$ and D7$_\alpha$,
whose number is given by the index of the Dirac operator on the
intersection 2-cycle $\Sigma^{\ff} \cap \Sigma^\alpha$ in the background of the world-volume
flux, cf.\ (\ref{index}).
In the example that the matter charged under the gauge group on D7$_\alpha$ transforms in the
(anti)fundamental representation, we verified explicitly that the charge of $T^\alpha$ and the
number of (anti)fundamentals take the right values to guarantee gauge invariance of the action
via the Green-Schwarz mechanism. If the gauge group on D7$_\alpha$ and the matter spectrum
allow for gaugino condensation (in which case we denoted D7$_\alpha$ as D7$_{\cc}$ before),
this implies automatically also the gauge invariance of the non-perturbative ADS superpotential
(\ref{wads}), present below the condensation-scale. Thus, our result complements the field theoretic
discussion of \cite{Achucarro:2006zf,Dudas:2006vc} by a more explicit embedding into a D-brane setup.

Furthermore, we also discussed the effects of higher curvature corrections to the D7-brane
action. The gauge kinetic function possibly receives a correction that depends on the
dilaton, cf.\ (\ref{fff}). We also argued that the charge of the axions is not modified
by the higher curvature corrections, thus leaving the mechanism described in the last paragraph
intact.

Our computation of the $D$-term potential
in Section 2 differs from the method used in \cite{Jockers:2005zy} in that
we determine the $D$-term potential directly from the dimensional reduction of the
bosonic DBI-action, whereas in \cite{Jockers:2005zy} the $D$-term potential is determined
indirectly from the fermionic terms and the standard supergravity relations.
The results, of course, agree with each other.

In our analysis, we concentrated our discussion on one flux-induced anomalous
$U(1)_{\ff}$ and one condensing gauge group $SU(N_{\cc})$. Clearly, more work is needed
to obtain a complete global model, which will be more involved. For example such a
model will generically feature several anomalous $U(1)$'s and the occurrence of
condensing gauge groups might be non-generic, see for instance \cite{Denef:2004dm}.
We expect that in such a model similar mechanisms
will be at work, although the structure
of effective field theories with several (pseudo-)anomalous $U(1)$'s and the
different types of cubic and mixed (as well as gravitational) anomalies
can be quite complicated and might also involve generalized Chern-Simons-terms
\cite{Anastasopoulos:2006cz}.

Another aspect that follows from the necessity of the charged matter
 fields $\Phi_I$, which was also discussed in \cite{Burgess:2003ic,Achucarro:2006zf,Dudas:2006vc},
is that the potential attains a much more complicated form,
 even in the case of only one K{\"a}hler modulus. Simply setting $\langle \Phi_{I}\rangle$
equal to zero renders  $W_{ADS}$ singular and is in general not a solution of the theory.
 Instead, the full scalar potential has to be minimized, also in the $\Phi_{I}$-directions
\cite{Achucarro:2006zf,Dudas:2006vc}, and including any tree-level contribution to the matter
superpotential. Only upon including all these effects
 can one decide whether the vev of $V_{D}$ is really non-zero or whether
 the matter fields relax to a vev that minimizes $V_{D}$ to zero preventing it from uplifting
the vacuum to a dS state. Unfortunately, the computation
 of tree-level matter superpotentials for intersecting D7-branes is rather complicated and
the result model-dependent. Also the presence of additional $B$-field moduli in the case of
$H^2_-(Y) \neq 0$ puts further constraints on the possibility of using $D$-term potentials
for uplifting, as minimizing the $D$-term potential with respect 
to these moduli tends to cancel the
uplifting $D$-term \cite{Jockers:2005zy,Lust:2006zg}.
An additional complication arises in the generic case of several K{\"a}hler moduli,
because then even the matter independent part of $V_D$ depends on both $T^{\cc}$ and
$T^{\ff}$ (and in general also on the other K{\"a}hler moduli due to the more complicated
form of the intersection numbers and hence the K{\"a}hler potential), as explained
 below (\ref{Dterm}).

 When the D7$_{\cc}$-stack has self-intersections and/or intersects O7-planes, one
encounters additional technical difficulties, as there would in general be
 matter fields in (anti-)symmetric tensor representations of $SU(N_{\cc})$.
A systematic analysis of the analogue of the ADS superpotential
does not seem to exist in the literature for this case.
For these technical reasons, we therefore restricted ourselves to  the case where the
D7$_{\cc}$-branes do not intersect
the O7-planes. Thus, in order to have a non-vanishing intersection of the D7$_{\cc}$- and
the D7$_{\ff}$-branes, we have to assume that also the latter
are not on top of the O7-planes (although they might intersect them). Our construction is
therefore prone to F-theory corrections, which we assume to be small due to sufficient proximity
of the D7-branes to the O7-planes. Taking into account F-theory and warping effects would be
an interesting, but challenging problem.

Another issue of the $D$-term uplifting with  $F$ fluxes  on D7-branes
is that these are quantized and also lack the analogue of the warp factor suppression of
the $\overline{D3}$ uplifting, as for D7-branes warped throats are not energetically
favored.\footnote{However, the D7-branes might extend into the throat, like recently discussed
for example in \cite{Baumann:2006th}, which would also lead to some suppression.}
This is sometimes criticized, as it seems to lead to a lack of sufficient tunability of
the uplift potential to a small value so as to make supersymmetry breaking effects small enough.
 In the large volume compactifications of
\cite{Balasubramanian:2005zx,Conlon:2005ki}, however, the $D$-terms are suppressed
 and can be very small, but in any case the charged matter contribution to the
$D$-terms has to be taken into account. A more substantial  analysis of these
issues is beyond the scope of the present paper.

Some interesting topics we have not considered in this paper include Euclidean instantons and
perturbative
corrections to the K{\"a}hler potential. Progress on the first issue has recently been made
in \cite{Ralph,Ibanez:2006da}.
The inclusion of perturbative corrections to the K{\"a}hler potential, like the ones discussed in
\cite{Becker:2002nn,Berg:2005ja,Antoniadis:1996vw}, would be interesting in view of the
proposal of \cite{Parameswaran:2006jh} which uses $D$-term potentials to uplift possible AdS minima
obtained from balancing effects in the $F$-term 
potential due to perturbative corrections to the
K{\"a}hler potential \cite{vonGersdorff:2005bf,Berg:2005yu}. 
A fully consistent discussion of the
uplift would
need to take into account the corrections to the K{\"a}hler 
potential in the $D$-term potential as well.

%%%%%%%%%%%%%%%%%%%%%%%%%%%%%%%%%%%%%%%%%%%%%

\begin{center}
{\bf Acknowledgements}
\end{center}
\vspace{-.3cm}

We would like to thank Marcus Berg, Massimo Bianchi, Ralph Blumenhagen, 
Gottfried Curio,  Jean-Pierre Derendinger,  Michael Dine,
Hans Jockers, Boris K{\"o}rs, Simon K{\"o}rs,  Jan Louis, Fernando Marchesano,
Hans-Peter Nilles, Mike Schulz,  and Stephan Stieberger for helpful discussions or email correspondence.
This work is supported in part by the European Community's Human
Potential Programme under contract MRTN-CT-2004-005104 `Constituents,
fundamental forces and symmetries of the universe'.
The work of M.~H. and M.~Z. is supported by the German Research Foundation (DFG) within
the Emmy-Noether-Program (grant numbers: HA 3448/3-1 and ZA 279/1-2). Moreover, M.H.\ and D.L.\ would
like to thank the KITP for hospitality during part of the project.
A.V.P.\ thanks the MPI and Arnold Sommerfeld Center in M{\"u}nchen for
hospitality during the completion of this work.
The work of A.V.P. is supported in part by the FWO - Vlaanderen, project
G.0235.05 and by the Federal Office for Scientific, Technical and
Cultural Affairs through the "Interuniversity Attraction Poles Programme
-- Belgian Science Policy" P5/27.

%%%%%%%%%%%%%%%%%%%%%%%%%%%%%%%%%%%%%%%%%%%%%%%%%%%%%%%%%%%%%%%%%%%%%%%%%%%%%%%

\begin{appendix}

%%%%%%%%%%%%%%%%%%%%%%%%%%%%%%%%%%%%%%%%%%%%%

\section{Example}
\label{example}

%%%%%%%%%%%%%%%%%%%%%%%%%%%%%%%%%%%%%%%%%%%%%

As toy model for the derivation of the $D$-term potential
we use the $\mathbb Z_2\times\mathbb Z_2$ orientifold
of a factorized $T_{(6)}$.
The metric $g_{(6)}$
of a factorized $T_{(6)}$ has block-diagonal form $g_{(6)}=(g_1,g_2,g_3)$ with
\begin{equation}\label{DTERMeq1}
g_\alpha=\left(
\begin{matrix}
(R_1^\alpha)^2&R_1^\alpha R_2^\alpha\cos\varphi^\alpha\\
R_1^\alpha R_2^\alpha\cos\varphi^\alpha&(R_2^\alpha)^2
\end{matrix}
\right),
\end{equation}
the metric of the respective sub-torus.

The geometrical moduli are defined as usual,
\begin{equation}\label{DTERMeq2}
\begin{split}
v^\alpha&:=R^\alpha_1R^\alpha_2\sin\varphi^\alpha,\\
\mathcal U^\alpha&:=\frac{R_2^\alpha}{R_1^\alpha}e^{i\varphi^\alpha},
\end{split}
\end{equation}
where $v^\alpha$ corresponds to the 2-cycle volumes.

Now consider a D7-brane wrapped on the four cycle $\Sigma^\alpha=T^\beta_{(2)}\times T^\gamma_{(2)}$
with $\alpha,\beta,\gamma$ all different. Furthermore, we assume that there is a non-trivial world-volume flux
present on this D7-brane.
Since the metric is explicitly known, one can directly calculate $\Gamma_\ff$ defined in (\ref{DBRANEeq16}) without invoking the supersymmetry condition (\ref{DBRANEeq3}) for the
D-brane:\footnote{As all the $(1,1)$-forms on $\Sigma^\alpha$ can be obtained by pullback from
the ambient space (due to the simplicity of the torus geometry),
there are no $\tilde f$ fluxes appearing here.}
\begin{equation}\label{DTERMeq4}
\Gamma_{\alpha}=\alpha^{\prime 2} \sqrt{(v^\beta)^2+(f^\beta)^2}\sqrt{(v^\gamma)^2+(f^\gamma)^2}.
\end{equation}
With
\begin{equation}\label{DTERMeq5}
\begin{split}
\Re\tilde\Gamma_{\alpha}&=\alpha^{\prime 2}(v^\beta v^\gamma-f^\beta f^\gamma),\\
\Im\tilde\Gamma_{\alpha}&=\alpha^{\prime 2}(v^\beta f^\gamma+v^\gamma f^\beta), %k
\end{split}
\end{equation}
we can rewrite $\Gamma_{\alpha}=|\tilde\Gamma_{\alpha}|$ which can be expanded as in (\ref{DBRANEeq17}) for small $\Im\tilde\Gamma_{\alpha}$:
\begin{equation}\label{DTERMeq6}
\Gamma_{\alpha}\approx \Re\tilde\Gamma_{\alpha}+\frac{1}{{\rm 2Re}\tilde\Gamma_{\alpha}}(\Im\tilde\Gamma_{\alpha})^2.
\end{equation}
It is well known, that in this setup supersymmetry is preserved if the following calibration condition between the 2-cycle volumes and the amount of 2-form flux on the D7-brane world-volume is fulfilled \cite{Berkooz:1996km}:
\begin{equation} \label{dtermcond}
\frac{v^\beta}{v^\gamma}=-\frac{f^\beta}{f^\gamma}.
\end{equation}
This condition leads to $\Im\tilde\Gamma_{\alpha}=0$.
Note that, for finite values of the
2-cycle volumes and world-volume fluxes, (\ref{dtermcond}) can only be fulfilled if $f^\beta$ and $f^\gamma$ differ in sign. Furthermore, in case of stabilization to a supersymmetric vacuum, this calibration condition leads to a partial fixing of the K{\"a}hler moduli \cite{Antoniadis:2004pp,Antoniadis:2005nu}.

The field-theoretical moduli are (cf.\ (\ref{kaehlermodsa}) and (\ref{kaehlermods})) \footnote{Up to  normalization, our moduli are defined as in \cite{Lust:2004cx}.}
\begin{equation}
\begin{split}
\Re T^\alpha&=\frac{1}{(2\pi)^5} e^{-\phi}v^\beta v^\gamma,\\\\
\Re S&=\frac{1}{(2\pi)^5} e^{-\phi}.
\end{split}
\end{equation}
The complex structure moduli do not need to be redefined $U^\alpha=\mathcal U^\alpha$.
Re-expressing $\tilde\Gamma_{\alpha}$ in this basis yields:
\begin{equation}
\begin{split}
\Re\tilde\Gamma_{\alpha}&=(2\pi)^5 \alpha^{\prime 2}e^\phi(T^\alpha_1-f^\beta f^\gamma S_1),\\
\Im\tilde\Gamma_{\alpha}&=\frac{\alpha^{\prime 2}}{(2\pi)^5}
\mathcal Ve^{-\phi}\left(\frac{f^\beta}{T^\gamma_1}+\frac{f^\gamma}{T^\beta_1}\right),
\end{split}
\end{equation}
where $\mathcal V=v^1v^2v^3$ is the volume and $T^\alpha_1$ and $S_1$
denote the real parts of the respective moduli.
Thus, the contribution to
the $D$-term scalar potential due to a single D7-brane and its orientifold image,
after transforming
to the Einstein-frame and omitting the term that is cancelled
after summing over all D-branes and O-planes, is given by (cf.\ (\ref{DBRANEeq17}))
\begin{equation} \label{exampledterm}
\frac12 \mu_7\,
e^{3\phi} \V^{-2}
\frac{1}{\Re\tilde\Gamma_\alpha}(\Im\tilde\Gamma_\alpha)^2=
\frac{1}{2g_{\alpha}^{-2}}\left(\frac{M_{\rm P}^2}{4\pi^2}
\left(\frac{(2 \pi)^{-2} f^\beta}{2 T^\gamma_1}
+\frac{(2 \pi)^{-2} f^\gamma}{2 T^\beta_1}\right)\right)^2,
\end{equation}
as expected from (\ref{DBRANEeq30}) with the tree-level K{\"a}hler potential
$K=-\sum_\beta\ln\left(T^\beta+\bar T^\beta\right)$. In deriving (\ref{exampledterm})
we made use of (\ref{mp}) and (\ref{DTERMeq5a}).

Since $\Sigma^\alpha$ is parameterized by $T^\alpha$, we observe that no $D$-term is
generated for the modulus parameterizing the 4-cycle the brane is wrapped on,
but rather for the moduli of the 4-cycles intersecting the wrapped cycle as discussed
before.\footnote{As mentioned earlier,
this changes in cases when the wrapped cycle has non-trivial self-intersections.}
%This is illustrated in figure \ref{DTERMfig1}.
%\begin{figure}
%\begin{center}
%\psfrag{t1}[bl]{\small $v_1$}
%\psfrag{t2}[bl]{\small $v_2$}
%\psfrag{t3}[bl]{\small $v_3$}
%\psfrag{T1}[b]{\small $T^3$}
%\psfrag{T2}[b]{\small $T^1$}
%\psfrag{T3}[b]{\small $T^2$}
%\includegraphics[scale=0.25]{t6.eps}
%\hspace{2cm}
%\includegraphics[scale=0.4]{toribranes.eps}
%
%\caption{Left: The 4-cycle $\Sigma^1$ is wrapped by a D7-brane with 2-form flux and thereby
%induces a $D$-term involving the K{\"a}hler moduli $T^2$ and $T^3$ of the intersecting
%4-cycles. Therefore, the issue of gauge invariance does not arise for gaugino condensation on
%the branes wrapping $\Sigma^1$.
%Right: The cycles $\Sigma^1$ and $\Sigma^3$ are wrapped by D7-branes with %world-volume fluxes.
%Now the modulus
%$T^1$ is charged under the $U(1)$ gauge field living on $T^3$ and vice versa. %Due to the
%intersection of the branes, bifundamental matter is present, leading to the %ADS superpotential.}
%\label{DTERMfig1}
%\end{center}
%\end{figure}
Therefore, for a single stack of D7-branes with 2-form flux, the question of
gauge invariance of the non-perturbative superpotential from gaugino
condensation on that same stack of branes does not arise.
If there is another stack of D7-branes on an
intersecting 4-cycle, bifundamental matter will
be present and the non-perturbative superpotential due to gaugino condensation on that
second brane stack
would be the Affleck-Dine-Seiberg superpotential which is naturally invariant under the
induced shift-symmetry as explained in the main text.

%%%%%%%%%%%%%%%%%%%%%%%%%%%%%%%%%%%%%%%%%%%%%%%%

\section{Alternative calculation of $\Gamma_\ff$}
\label{alternative}

%%%%%%%%%%%%%%%%%%%%%%%%%%%%%%%%%%%%%%%%%%%%%%%%

In this appendix we give an alternative derivation of (\ref{DBRANEeq40})
which follows in spirit more closely the method of \cite{Villadoro:2006ia} in that
it does not take (\ref{DBRANEeq3}) as a starting point.
However, we still require that the world volume flux is of type $(1,1)$,
i.e.\ (\ref{DBRANEeq4}) holds. Furthermore, we neglect the higher derivative
corrections to the DBI-action in the following.

First, we note that there is the following relation between
quantities in real and complex coordinates
\begin{equation}
(\io g)_{mn} + \F_{mn}= \left( \begin{array}{cc}
                          0 & (\io g)_{i \bar \jmath} - \F_{i \bar \jmath} \\
                          (\io g)_{i \bar \jmath} + \F_{i \bar \jmath} & 0
                         \end{array}
         \right)\ .
\end{equation}
Thus
\begin{eqnarray}
\det ((\io g)_{mn} + \F_{mn}) &=& \det ((\io g)_{i \bar \jmath} - \F_{i \bar \jmath})
\det ((\io g)_{i \bar \jmath} + \F_{i \bar \jmath}) \non
&=& \det (-i (\io J)_{i \bar \jmath} - \F_{i \bar \jmath})
\det (-i (\io J)_{i \bar \jmath} + \F_{i \bar \jmath}) \non
&=& \det ((\io J)_{i \bar \jmath} - i \F_{i \bar \jmath})
\det ((\io J)_{i \bar \jmath} + i \F_{i \bar \jmath})\ ,
\end{eqnarray}
which implies
\begin{eqnarray} \label{sqrtdet}
\sqrt{\det ((\io g)_{mn} + \F_{mn})}
&=& \sqrt{\det ((\io J)_{i \bar \jmath} - i \F_{i \bar \jmath})
\det ((\io J)_{i \bar \jmath} + i \F_{i \bar \jmath})} \\
&=&
\sqrt{\Big(\frac12 \epsilon^{ik} \epsilon^{\bar \jmath \bar l}
((\io J)_{i \bar \jmath} (\io J)_{k \bar l}
- \F_{i \bar \jmath} \F_{k \bar l})\Big)^2
+ \Big(\epsilon^{ik} \epsilon^{\bar \jmath \bar l} (\io J)_{i \bar \jmath}
\F_{k \bar l}\Big)^2}\ , \nonumber
\end{eqnarray}
where our epsilon-symbol takes values $0$ and $\pm 1$.

Next, consider two harmonic $(1,1)$-forms
$\omega^{(1)}$ and $\omega^{(2)}$ on the 4-cycle $\Sigma^{\ff}$ that is
wrapped by the D7-brane.
It follows that $\omega^{(1)} \wedge \omega^{(2)} \in H^{(2,2)}(\Sigma^{\ff})$ and
\begin{eqnarray}
\star (\omega^{(1)} \wedge \omega^{(2)}) &=& \star \left(
\omega^{(1)}_{i \bar \jmath}
\omega^{(2)}_{k \bar l}\, d z^i \wedge d \bar z^{\bar \jmath} \wedge
d z^k \wedge d \bar z^{\bar l} \right) \non
&=& \star \left( - \omega^{(1)}_{i \bar \jmath}
\omega^{(2)}_{k \bar l}\, d z^i \wedge d z^k \wedge d \bar z^{\bar \jmath}
\wedge d \bar z^{\bar l} \right) \non
&=& -\frac14 \omega^{(1)}_{i \bar \jmath} \omega^{(2)}_{k \bar l}
\epsilon^{ik} \epsilon^{\bar \jmath \bar l}\sqrt{\det(\io g)}^{-1} \quad \in H^{(0,0)}(\Sigma^{\ff})\ ,
\end{eqnarray}
where with $\sqrt{\det(\io g)}$ we always mean $\sqrt{\det((\io g)_{mn})}=\det((\io g)_{i \jmath}$.
As any harmonic function on a compact Riemannian manifold is constant, we
infer that $\omega^{(1)}_{i \bar \jmath} \omega^{(2)}_{k \bar l}
\epsilon^{ik} \epsilon^{\bar \jmath \bar l}\sqrt{\det(\io g)}^{-1}$ is constant.
On the other hand, one has
\begin{equation}
\int_{\Sigma^{\ff}} \omega^{(1)} \wedge \omega^{(2)} =
-\int_{\Sigma^{\ff}} \omega^{(1)}_{i \bar \jmath} \omega^{(2)}_{k \bar l}
\epsilon^{ik} \epsilon^{\bar \jmath \bar l} d^4 z
= - \omega^{(1)}_{i \bar \jmath} \omega^{(2)}_{k \bar l}
\epsilon^{ik} \epsilon^{\bar \jmath \bar l} \sqrt{\det(\io g)}^{-1} \V_{\ff}\ ,
\end{equation}
where $\V_{\ff}$ is the volume of the 4-cycle $\Sigma^{\ff}$. Thus one concludes
\begin{equation}
\omega^{(1)}_{i \bar \jmath} \omega^{(2)}_{k \bar l}
\epsilon^{ik} \epsilon^{\bar \jmath \bar l} = - \sqrt{\det(\io g)} \V_{\ff}^{-1}
\int_{\Sigma^{\ff}} \omega^{(1)} \wedge \omega^{(2)}\ .
\end{equation}
Plugging this into (\ref{sqrtdet}), we derive
\begin{equation}
\sqrt{\det ((\io g) + \F)}
= \V_{\ff}^{-1} \sqrt{\det(\io g)}
\sqrt{\frac14 \left(\int_{\Sigma^{\ff}} \io J \wedge \io J - \int_{\Sigma^{\ff}} \F \wedge \F \right)^2
+ \left(\int_{\Sigma^{\ff}} \io J \wedge \F \right)^2}\ .
\end{equation}
Using this in (\ref{DBRANEeq16}), we again arrive at
\begin{equation}
\Gamma_\ff = \sqrt{\frac14 \left(\int_{\Sigma^{\ff}} \io J \wedge \io J
- \int_{\Sigma^{\ff}} \F \wedge \F \right)^2
+ \left(\int_{\Sigma^{\ff}} \io J \wedge \F \right)^2}\ ,
\end{equation}
which coincides with (\ref{DBRANEeq40}).

%%%%%%%%%%%%%%%%%%%%%%%%%%%%%%%%%%%%%%%%%%%%%%%%%%%%%%%%%%%%%

\section{Gauge kinetic function}
\label{gaugekin}

%%%%%%%%%%%%%%%%%%%%%%%%%%%%%%%%%%%%%%%%%%%%%%%%%%%%%%%%%%%%%

The imaginary part of the gauge kinetic function is due to the
following terms of the Chern-Simons action:
\begin{equation}
\begin{split}
&-\frac12 \mu_7\int_{\mathcal W}\iota^*C_4 \wedge\mathcal F\wedge \mathcal F,\\
&-\frac{1}{4!} \mu_7\int_{\mathcal W}C_0 \wedge\mathcal F\wedge
\mathcal F\wedge\mathcal F\wedge \mathcal F.
\end{split}
\end{equation}
In the first term, $C_4$ lives purely internal on $\Sigma^{\ff}$ while
both $\mathcal F$ denote external field strengths with indices
along $\mathcal M_4$:
\begin{equation}
-\frac12 \mu_7(2\pi\alpha')^2 \int_{\mathcal M_4} \Big( \int_{\Sigma^{\ff}}
\iota^*C_4 \Big)
F\wedge F =-\frac12 \mu_7(2\pi)^2 \alpha^{\prime 4}
\int_{\mathcal M_4}a^\ff\, F\wedge F,
\end{equation}
where we defined $a^\ff=\alpha^{\prime -2}
\int_{\Sigma^{\ff}}\iota^*C_4$.

In the second term two $\mathcal F$ are assumed to live purely on
$\mathcal M_4$ while the other two $\mathcal F$ are assumed to have
internal indices:
\begin{equation}\label{APPCeq1}
-\frac12 \mu_7 (2\pi\alpha')^2\int_{\Sigma^{\ff}}\mathcal F\wedge
\mathcal F\int_{\mathcal M_4}
\frac12 C_0 F\wedge F=-\frac12 \mu_7 (2\pi)^2 \alpha^{\prime 4}
\mathfrak{f}_\ff \int_{\mathcal M_4}
C_0 F\wedge F,
\end{equation}
where we expanded the 2-form flux as in (\ref{DBRANEeq6}) and $\mathfrak{f}_\ff$
was defined
in (\ref{fhat}).

Thus, the imaginary part of the gauge kinetic function
$\Im f^\ff$ is given by
\begin{equation}
\Im f^\ff=\mu_7(2\pi)^2 \alpha^{\prime 4}
\left(a^\ff+\mathfrak{f}_\ff C_0 \right)= \frac{1}{(2 \pi)^5}
\left(a^\ff+\mathfrak{f}_\ff C_0 \right)\ ,
\end{equation}
where we made use of (\ref{mu7b}) in the second equality.
Comparing with the real part (\ref{realpart}) and demanding holomorphicity
of the gauge kinetic functions, we arrive at (\ref{kaehlermods}).

Let us now take the curvature corrections into account.
The corrections to the CS-action were already given in (\ref{CScorr}).
Since only terms of the CS-action contribute to the imaginary part
of the gauge kinetic function which possess two external field-strengths, we see that
only the following additional terms arise:
\begin{equation}
-\frac{1}{2}\frac{1}{48} \mu_7\int_{\mathcal W}C_0 \wedge\mathcal F\wedge
\mathcal F\wedge \Big(-p_1(T(\Sigma^{\ff})) + p_1(N(\Sigma^{\ff}))\Big)\ ,
\end{equation}
where $\mathcal F$ is an external field strength.
Hence, similar to (\ref{APPCeq1}) we obtain
\begin{equation}
-\frac{1}{48} \mu_7 (2\pi\alpha')^2\int_{\Sigma^{\ff}} \Big( - p_1(T(\Sigma^{\ff}))
+p_1(N(\Sigma^{\ff}))
\Big) \int_{\mathcal M_4}
\frac12 C_0 F\wedge F\ .
\end{equation}
Thus the curvature corrected imaginary part of the gauge kinetic function is given by
\begin{equation}
\Im f^\ff=\frac{1}{(2 \pi)^5}
\left(a^\ff+\hat{\mathfrak{f}}_\ff C_0 \right)\ ,
\end{equation}
where $\hat{\mathfrak{f}}_\ff$ was defined in (\ref{hatfrakf}).

\end{appendix}

%%%%%%%%%%%%%%%%%%%%%%%%%%%%%%%%%%%%%%%%%%%%%%%%%%%%%%%%%%%%%%%%%%%%%%%%%%%%%%%%

\end{document}